\documentclass{aa}  
\usepackage{natbib}
\usepackage[pdftex]{graphicx}
\bibpunct{(}{)}{;}{a}{}{,}
\usepackage{graphicx}
\usepackage{txfonts}
\begin{document}
\def\deg{\hbox{$^\circ$}}
\def\arcmin{\hbox{$^\prime$}}
   \title{A Survey of the Polarized Emission from the Galactic Plane at
1420~MHz with Arcminute Angular Resolution}

%   \subtitle{}

\author{T.L. Landecker \inst{1} \and W. Reich \inst{2} \and R.I. Reid
\inst{1,3} \and P. Reich \inst{2} \and M. Wolleben \inst{1,2} \and
R. Kothes \inst {1,4} \and B. Uyan{\i}ker \inst{1,5} \and A.D. Gray
\inst{1} \and D. Del Rizzo \inst{1} \and E. F{\"u}rst \inst{2} 
\and A.R. Taylor \inst{4} \and R. Wielebinski \inst{2} }

\titlerunning{Polarization survey at 1420 MHz}
\authorrunning{Landecker et al.}

   \offprints{T.L. Landecker}

\institute{National Research Council of Canada,
              Herzberg Institute of Astrophysics,
              Dominion Radio Astrophysical Observatory,
              P.O. Box 248, Penticton, British Columbia,
              V2A 6J9, Canada\\
\and
              Max-Planck-Institut f\"ur Radioastronomie,
              Auf dem H\"ugel 69, 53121 Bonn, Germany\\
\and
              Present address: National Radio Astronomy Observatory,
              520 Edgemont Road, Charlottesville, Virginia 22903-2475, 
              USA \\
\and
              Department of Physics and Astronomy, University of
              Calgary, 2500 University Drive N.W., Calgary, AB,
              Canada\\
\and
              Present address: 35/3737 Gellatly Road, Westbank, British
	      Columbia, V2T 2W8, Canada\\
}
   \date{Received ; accepted }

% \abstract{}{}{}{}{} 
% 5 {} token are mandatory
   \abstract
   {Observations of polarized emission are a significant source of
   information on the magnetic field that pervades the Interstellar
   Medium of the Galaxy. Despite the acknowledged importance of
   magnetic field in interstellar processes, our knowledge of field
   configurations on all scales is seriously limited.}
   {This paper describes an extensive survey of polarized Galactic
   emission at 1.4~GHz that provides data with arcminute resolution
   and complete coverage of all structures from the broadest angular
   scales to the resolution limit, giving information on the
   magneto-ionic medium over a wide range of interstellar
   environments. }
   {Data from the DRAO Synthesis Telescope, the Effelsberg 100-m
   Telescope, and the DRAO 26-m Telescope have been combined. Angular
   resolution is ${\sim}1'$ and the survey extends from
   ${\ell}={66^{\circ}}$ to ${\ell}={175^{\circ}}$ over a range
   ${-3^{\circ}} < b < {5^{\circ}}$ along the northern Galactic plane,
   with a high-latitude extension from ${\ell}={101^{\circ}}$ to
   ${\ell}={116^{\circ}}$ up to ${b}={17\fdg5}$. This is the first
   extensive polarization survey to present aperture-synthesis data
   combined with data from single antennas, and the techniques
   developed to achieve this combination are described.}
   {The appearance of the extended polarized emission at 1.4~GHz is
   dominated by Faraday rotation along the propagation path, and the
   diffuse polarized sky bears little resemblance to the
   total-intensity sky.  There is extensive depolarization, arising
   from vector averaging on long lines of sight, from \ion{H}{ii}
   regions, and from diffuse ionized gas seen in H$\alpha$ images.
   Preliminary interpretation is presented of selected polarization
   features on scales from parsecs (the planetary nebula Sh 2-216) to
   hundreds of parsecs (a superbubble GSH~166$-$01$-$17), to
   kiloparsecs (polarized emission in the direction of Cygnus~X).}
   {}


\maketitle
%
%________________________________________________________________

\section{Introduction}
\label{sec:intro}

The detection of linear polarization in the Galactic radio emission
\citep{west62,wiel62} provided crucial evidence in establishing the
synchrotron mechanism as the source of the emission. Assumption of
equipartition between relativistic particles and magnetic field then
led to estimates of the field strength of a few $\mu$G \citep{beck01},
and the best estimates today do not differ substantially
\citep{sun08}. The magnetic field in the Galaxy is a significant
reservoir of energy, and the field is likely to play an important role
in interstellar processes. Nevertheless, more than four decades after
the first detection our knowledge of the field configuration on global
and local scales is still quite limited.

The apparent promise of polarization observations has proved difficult
to translate into hard information on field configurations. First, the
observed fractional polarization is generally far below the 70\%
theoretical maximum.  The radiation is optically thin and
superposition of emission contributions along the line of sight can
``depolarize'' the signal through vector averaging. Second, Faraday
rotation operates whenever the signal propagates through a magnetized
thermal plasma.  Emission and Faraday rotation often occur in the same
region, and a wide variety of depolarization effects occur
\citep{burn66,soko98}.  Furthermore, a typical radio telescope
operating at low radio frequencies proves to be more sensitive to the
Faraday rotation of a plasma region than to its bremsstrahlung (as 
discussed in Section~\ref{subsec:small-scale}).
The net result of all these effects is that the
polarized sky rarely resembles the total-intensity sky and
interpretation of polarization images is seldom simple.

Angular resolution of early polarization surveys with single antennas
was poor.  A comprehensive set of surveys of the northern sky was
published by \citet{brou76} based on well calibrated (but
undersampled) observations at four frequencies between 408 MHz and
1411 MHz made in the 1960s with the Dwingeloo 25-m Telescope; angular
resolution ranged from 2$^{\circ}$ to 36\arcmin. Here the subject
rested for many years until it was revived using the Effelsberg 100-m
Telescope in the late 1980s. Extensive surveys made with that
telescope had resolutions of 4\farcm3 at 2695~MHz
\citep{junk87,dunc99} and 9\farcm4 at 1410~MHz
\citep{uyan98,uyan99,reic04}. A wide-area survey of the Southern
Galactic plane was made with the Parkes Telescope at 2417~MHz with
resolution 10\farcm4\, \citep{dunc97}. The Northern Galactic plane
was surveyed with the Urumqi Telescope with resolution 9\farcm5 at
4900~MHz \citep{sun07,gao10}. The WMAP data at 23~GHz and higher frequencies
\citep{hins09} cover the entire sky at an angular resolution smaller than
$1^{\circ}$ and, on simple lines of sight, show almost the intrinsic
polarization characteristics because Faraday rotation at these
frequencies is extremely low.

A substantial increase in angular resolution has been provided by
aperture-synthesis telescopes, and observations of the Galactic
polarized emission with angular resolutions from one to a few
arcminutes have been made in recent years
\citep{wier93,gray98,gray99,have00,gaen01,uyan02,uyan03,have03a,have03b,
have03c,have06a}.  These observations have revealed much about Faraday
rotation in the ISM, but interpretation has been hampered by the lack
of information on the biggest structures{\footnote{ Single-antenna
data may also suffer from this problem -- see
Section~\ref{subsec:eff_process} and \citet{reic06}}}.  

An interferometer observes the sky through a spatial high-pass filter,
so the zero levels for $Q$ and $U$ are lost. Both polarized intensity,
${\rm{PI}}={\sqrt{Q^{2}+U^{2}}}$, and polarization angle,
${\rm{PA}}={{0.5}\,{\rm{tan}}^{-1}\left(\frac{U}{Q}\right)}$, change
in a very non-linear fashion in response to errors in zero levels.
The most serious effect is on angle. Consider an observation of a
region several tens of beamwidths in extent comprising broad
structure, whose polarization angle, $A$, varies only slowly across
the area, and fine structure, contributing rapid changes in PA that
add vectorially to $A$. An interferometer can measure only the rapid
angle changes, and, if the broad structure is not measured, the
apparent PA will vary over a large range, whereas the true
distribution of angle is much narrower and is centred on $A$.
Examples of this effect are illustrated in \citet{reic06},
\citet{sun07}, and \citet{gao10}. An example in the present data is
discussed in Section~\ref{subsec:w543}. If broad structure is missing
then conclusions about PA, and hence rotation measure (RM), are prone
to serious error. In contrast, the work described here is the first
extensive polarization survey to incorporate single-antenna data with
aperture-synthesis data.

In this paper we present a survey of the polarized emission at
1420~MHz covering 1060 square degrees of the northern Galactic plane
with an angular resolution of $\sim$1~arcminute.  With $1.5 \times
10^7$ independent data points this is the largest polarization survey
published to date. The survey combines data from the DRAO Synthesis
Telescope, the Effelsberg 100-m Telescope, and the DRAO 26-m
Telescope, and we discuss the techniques that we have developed to
combine these datasets. Data from two single-antenna telescopes, not
just one, were needed to correctly represent broad structure for reasons
discussed below (Section~\ref{subsec:eff_process}).

The new polarization dataset forms part of the Canadian Galactic Plane
Survey (the CGPS, described by Taylor et al. 2003). The scientific
goal of the CGPS is the study of the Galactic ``ecosystem'', the
interplay between the various constituents of the ISM, their role in
star formation, the impact of stars on their environments, and the
interaction of Galaxy-wide phenomena such as density waves with ISM
constituents. The CGPS comprises surveys{\footnote{The CGPS database
is accessible to the astronomy community at
{\tt{http://www3.cadc-ccda.hia-iha.nrc-cnrc.gc.ca/cgps/}}}} of the
atomic hydrogen (the 21-cm \ion{H}{i} line), the ionized gas (seen in radio
continuum), the molecular gas (the lines of CO near 115~GHz -
\citep{heye98}), the dust (IRAS data reprocessed with high resolution -
\citep{cao97,kert00}), and, relevant here, the relativistic component
and the magnetic field traced by continuum observations at 408 MHz
(total intensity only) and 1420 MHz (Stokes parameters $I$, $Q$, and
$U$).  The survey described here maps the magneto-ionic component of
the ISM with unprecedented detail and precision.

\section{Telescopes and signal processing equipment}
\label{sec:telescopes}

\subsection{The DRAO Synthesis Telescope}
\label{subsec:drao_st}

A detailed description of the DRAO Synthesis Telescope can be found in
\citet{land00}; only an outline is given here, with emphasis on
changes since that paper was published and on polarization properties.
Telescope characteristics relevant to this survey are summarized in
Table~\ref{tabl-st}.

\begin{table}
\caption[]{DRAO Synthesis Telescope survey characteristics}
\label{tabl-st}
\begin{center}
\begin{tabular}{ll}
\hline
Coverage                     & ${66^{\circ}} < {\ell} < {175^{\circ}}, 
                               {-3^{\circ}} < {b} < {5^{\circ}}$; \\
                             & ${101^{\circ}} < {\ell} < {116^{\circ}}, 
                               {5.0^{\circ}} < {b} < {17\fdg5}$ \\
Individual field size        & 107\farcm2~diameter to 50\% \\
                             & 150\farcm0~diameter to 25.7\% \\
Field used in mosaicking     & 150\arcmin~diameter \\
Field centre spacing         & 112\arcmin\ on triangular grid \\
Observation dates            & 1995.3 to 2004.4 \\
Survey area                  & 1060 square degrees \\
Observing rate               & 150 square degrees per year \\
Bandwidth                    & 30\,MHz in four bands of \\
                             & 7.5\,MHz each \\
Centre frequencies           & 1407.2, 1414.1, 1427.7,\\
                             & and 1434.6\,MHz \\
Angular resolution           & $58'' \times 58''$\,cosec\thinspace$\delta$ \\
System temperature           & 60~K, 1995.3 to 2003.4; \\
                             & 45~K, 2003.4 to 2004.4 \\
Sensitivity (mosaicked data) & 0.30 mJy/beam rms 1995.3 to 2003.4 \\
                             & ${76}\thinspace{\rm{sin{\thinspace}(declination)}}$~mK; \\
                             & 0.23 mJy/beam rms 2003.4 to 2004.4 \\
                             & ${58}\thinspace{\rm{sin{\thinspace}(declination)}}$~mK \\
\hline
\end{tabular}
\end{center}
\end{table}

The telescope consists of seven antennas, of diameter $\sim$9 m. The
antennas receive right-hand and left-hand circular polarization (RHCP
and LHCP) at 1420~MHz; the maximum baseline of 617~m gives an angular
resolution of ${\sim}1'$ at that frequency. A telescope attribute that
strongly influences polarimetry is the slight difference in
construction of the seven antennas. Two antennas have diameter
${d}={9.14}~\rm{m}$ and focal length ${f}={3.81}~\rm{m}$
(${f/d}=0.42$) while the remainder have ${d}={8.53}~\rm{m}$ and focal
length ${f}={3.66}~\rm{m}$ (${f/d}=0.43$). The symmetrical reflectors
have prime-focus feeds supported by struts whose construction differs
slightly on the various antennas. These seemingly minor differences
have a considerable effect on polarization properties and special
procedures have been developed to correct images of polarized emission
for instrumental polarization (see Section~\ref{subsec:st_instpol}).

For continuum imaging at 1420~MHz the telescope has four frequency
bands, each of width ${{\Delta}f}={7.5}$~MHz, two below and two above
the frequency of the \ion{H}{i} line. In operation the telescope is
tuned to the \ion{H}{i} centre frequency; because of changes in the
relative velocity of telescope and source the centre frequency
changes (for example due to the Earth's rotation and orbital motion).
The image formation algorithms use the exact centre frequency at which
a data sample was observed (different for each band). For the CGPS
observations the central velocity to which the receiver is tuned
varies from $v_{lsr}=-40$~km~s$^{-1}$ at ${\ell}={66^{\circ}}$ to
$v_{lsr}=-60$~km~s$^{-1}$ at ${\ell}={175^{\circ}}$. This variation
corresponds to a frequency change of about 90~kHz, 1.3\% of the width
of an individual band and an inconsequential fraction of the centre
frequency.  The nominal centres of the four continuum bands lie at
$\pm$6.25~MHz and $\pm$13.75~MHz from the central tuning frequency
but small bandshape errors mean that the effective centre frequencies
deviate slightly from these values. The deviations have been measured
and are taken into account; the effective centre frequencies for the
four bands are listed in Table~\ref{tabl-st}. Velocity corrected
effective centre frequencies are preserved in data headers and are
used, for example, in calculations of RM.

The DRAO Synthesis Telescope is characterized by good sensitivity to
extended structure and a very thorough sampling of the {\it{u-v}}
plane. The shortest baseline is 12.9~m, 61 wavelengths at 1420~MHz,
nominally giving the telescope sensitivity to structures at least as
large as $\sim$40\arcmin. The {\it{u-v}} sampling interval is
${L}={4.29}~{\rm{m}}$, about half the antenna diameter. Consequently,
the first grating response is at an angular radius where the antenna
response is very low and grating responses to objects within the field
of view lie well outside the field.

\subsection{The Effelsberg 100-m Telescope}
\label{subsec:eff_100m}

The Effelsberg 100-m telescope \citep{hach73} is a fully steerable
high precision antenna with surface deviations of about 0.5~mm (rms)
that is based on the concept of homologous elevation-dependent
deformations. L-band observations are always made from the prime
focus. In this frequency band elevation-dependent gain variations can
be ignored.

A detailed measurement of the antenna pattern at $\lambda$21\ cm was
made by \citet{kalb80}. A feed with a slightly different taper was
used later, when \citet{uyan98} measured an area of $1\fdg1 \times
1\fdg1$ around 3C~123 covering the main beam and the first and
second sidelobes. The shape of the main beam can be very well
approximated by a circular Gaussian with a HPBW of
$9\farcm35$. Sidelobes are enhanced in the direction of the four
subreflector support struts and the first sidelobes have peak responses
of about $-$20~dB. The conversion between main beam brightness
temperatures $\rm T_{B}$ and Jy/beam~area is 2.12~K/Jy. The L-band
receiving system has two channels with 
cooled HEMT amplifiers for LHCP and RHCP covering 1.29 to 1.72~GHz, and has a
system noise of about 26~K. Because of interference the
frequency band for continuum observations has to be close to 1.4~GHz with a
typical bandwidth of 20~MHz, avoiding Galactic
\ion{H}{i} emission. An IF correlation polarimeter
provides two total power channels and two correlated channels
proportional to Stokes $U$ and $Q$. Further details of the receiving
system are given by \citet{uyan98}. In 2001 a second IF-polarimeter
was added allowing simultaneous observations with centre frequencies
at 1395~MHz and 1408~MHz and a bandwidth of 14~MHz each. 
A new IF polarimeter was installed in 2002, offering eight 4-MHz channels, normally centred at 1402~MHz. Telescope characteristics relevant to this survey are
summarized in Table~\ref{tabl-100m}.

\begin{table}
\caption[]{Effelsberg 100-m Telescope survey characteristics}
\label{tabl-100m}
\begin{center}
\begin{tabular}{ll}
\hline
Coverage                     & ${25^{\circ}} < {\ell} < {230^{\circ}},$\\ 
                             & ${-20^{\circ}} < {b} < {20^{\circ}}$; \\
Observation dates            & 1994 to 2005 \\
Bandwidth $(B)^1$                & 20 MHz \\
Frequency $(F)^1$                & 1400 MHz \\
Angular resolution           & 9\farcm35 \\
System temperature           & 26~K \\
Sensitivity (rms noise $U$,$Q$)   & 8~mK \\
\hline
\end{tabular}
\end{center}
$^1$~Occasional interference required decreases in $B$, which never fell
below 10 MHz, and/or increases in $F$, which never rose above 1410~MHz.
\end{table}

\subsection{The DRAO 26-m Telescope}
\label{subsec:drao_26m}

The DRAO 26-Telescope is an equatorially mounted, axially symmetric
reflector. At 1.4 GHz the aperture efficiency is $\sim$53\%. Pointing
accuracy is $\sim$1$'$. Data for the present work were drawn from the
survey of \citet{woll06} and the telescope and receiver are described
in full in that paper; relevant telescope characteristics are
summarized in Table~\ref{tabl-26m}.

\begin{table}
\caption[]{DRAO 26-m Telescope survey characteristics}
\label{tabl-26m}
\begin{center}
\begin{tabular}{ll}
\hline
Coverage                     & Right Ascension: $0^h$ to $24^h$, \\ 
                             & Declination: ${-29^{\circ}}$ to ${90^{\circ}}$ \\
Observation dates            & 2002.9 to 2003.5 and 2004.5 to 2005.3 \\
Bandwidth                    & 12 MHz \\
Frequency                    & 1410 MHz \\
Angular resolution           & 36\farcm0 \\
System temperature           & 125~K \\
Sensitivity (rms noise $U$,$Q$) & 12~mK \\
\hline
\end{tabular}
\end{center}
\end{table}

The two hands of circular polarization were generated from linear
outputs of the prime-focus feed using a coaxial quadrature hybrid.
The beamwidth was 36$'$ at 1.4~GHz. The receiver was uncooled, and
yielded a system temperature of $\sim$125~K.  The RHCP and LHCP
outputs were processed in an IF analog polarimeter, of the type used
for the Effelsberg observations, which generated all polarization
products.

\section{Observation and processing of Synthesis Telescope data}
\label{sec:processing}

\subsection{Observations}
\label{subsec:st_obs}

The DRAO Synthesis Telescope has only seven antennas but samples the
{\it{u-v}} plane thoroughly using multiple array configurations; as a
consequence the mapping speed is slow, only 150 square degrees per
year.  The observation period for the present survey was long, from
1995 April to 2004 March.  Observations were made day and night,
although solar interference and increased Faraday rotation in the
ionosphere make daytime less than ideal for polarimetry.  Unavoidably,
over the long period, some telescope parameters changed, but, on the
positive side, image processing techniques were improved. The
observations began just before a minimum of solar activity (at
$\sim$1996.5), through a maximum at $\sim$2001, and well into a second
minimum, with the attendant changes in ionospheric conditions. For all
these reasons the data product is less uniform than one observed over
a shorter time span.

Full details of the observation technique for the CGPS can be found in
\citet{tayl03}.  Details relevant to the polarization survey can be
found in Table 1. The latitude coverage of the survey,
${-3^{\circ}}<{b}<{5^{\circ}}$, was biased north of the nominal
Galactic plane because of the warp of the northern Galactic disk.

\subsection{Calibration}
\label{subsec:st_cal}

The telescope was calibrated using small-diameter extragalactic
sources, with preference given to 3C~147 and 3C~295, known to be
unpolarized, but 3C~48 was also used \citep{smeg97}. Observations were
interspersed with calibrations, usually at intervals of
12~hours. Assumed flux densities are consistent with the scale of
\citet{baar77} -- see \citet{land00} for details. After
calibration the correlator outputs are known in units of flux density;
this applies to polarization outputs as well as total-intensity
outputs. The PI calibration of the survey is
therefore tied to the well established flux density scale.
PA was calibrated from observations of the
polarized source 3C~286, made at intervals of four days
\citep{tayl03}.

\subsection{Image processing}
\label{subsec:image_proc}

Image formation generally followed the conventional practices of
aperture-synthesis technique; processing specific to the CGPS is
described in \citet{tayl03}. The modified {\small{CLEAN}} algorithm
developed by \citet{stee84} works well on images with extended
structure and was used exclusively.  The usual next step,
self-calibration, is ineffective on polarization images because they
contain very little flux. Because antenna phase and amplitude gains
are the same for all cross-correlations, gain solutions derived
from self-calibration of RHCP and LHCP total-intensity images could be
applied to the cross-hand products that were used to make polarization
images.

After these initial procedures some artefacts remained in the images,
centred on bright sources inside and outside the primary beam.
Because of the wide field of view and small differences between the
individual antennas of the array, complex gain errors occur at large
distances from the field centre. The resulting artefacts affect $I$
data but are more pronounced in $Q$ and $U$ images because the
instrumental polarization rises rapidly with distance from the field
centre (see Section~\ref{subsec:st_instpol}). The artefacts are not
easily removable by standard self-calibration techniques because the
assumption of uniform complex gains over the primary beam is
incorrect. They are removed by the procedure called {\small{MODCAL}}
\citep{will99} which is, in principle, a position-dependent self
calibration.

Initially this procedure was used for strong sources as close as
40\arcmin\ from the field centre. Later in the progress of the survey
the instrumental polarization correction described in
Section~\ref{subsec:st_instpol} was applied and artefacts around
strong sources were reduced to the point where use of {\small{MODCAL}} was
seldom required.

\subsection{Solar and terrestrial interference and emission from the ground}
\label{subsec:solar}

The antenna sidelobes are highly polarized and sources outside the
primary beam can produce strong spurious polarized signals even if
their emission is not inherently polarized.  The Sun is such a source:
it traverses the antenna sidelobes whenever it is above the horizon,
contributing $\sim$1~K to antenna temperature.  Conversion of $I$ to
$Q$ or $U$ can be as high as 50\% in the far sidelobes \citep{ng05}
so the Sun can contribute a polarized signal up to 500~mK. However,
because of the Sun's large extent its effects are usually confined to
shorter baselines. Spurious solar signals were removed to a
satisfactory level by making images centred on the Sun's position and
removing the response from the visibilities. Use of {\small{MODCAL}}
to adjust phases in the direction of the Sun improved the removal
process.

Terrestrial interference, which is always polarized, is another source
of spurious polarization. Interference may come from any direction,
but, because the telescope is an east-west interferometer, persistent
terrestrial interference appears in images as rings centred at the
north celestial pole.  Making a 640\arcmin\ wide image of the north
polar cap and subtracting its Fourier transform from the visibilities
proved to be an effective way of removing it. Again, {\small{MODCAL}}
improved the removal.

Radiation from the ground appears as polarized emission because of the
conversion of $I$ into $Q$ and $U$ by the sidelobes. A patch of ground
can radiate the same signal into two adjacent antennas when they are
close together, producing a strong correlated signal in polarization
channels. This effect is more difficult to remove, and polarization
data gathered at the shortest baseline (${\sim}$12.9~m) were mostly
unusable for polarization imaging (while being quite usable for
imaging of $I$ and \ion{H}{i}). Data from the Effelsberg Telescope
were used to provide these spatial frequencies for polarization
imaging.

When a reasonably good image can be formed of an interfering source,
subtracting its Fourier transform from the visibilities is much better
than completely removing (``flagging'') the affected data, but there
is inevitably some confusion with real structure on the sky at those
spatial frequencies.  However, in some particularly bad cases the data
for affected interferometer spacings were simply flagged and removed.

\subsection{Correction for instrumental polarization}
\label{subsec:st_instpol}

Two levels of correction for instrumental polarization were required.
The feed and associated waveguide components do not provide perfect
LHCP and RHCP and there is coupling between the two polarized
outputs. The regular telescope calibration (see Section~\ref{subsec:st_cal})
corrected for these effects, which apply to the whole field of
view. The calibration technique is described in \citet{smeg97}.  After
this calibration the spurious polarization at field centre was
nominally zero with an accuracy estimated to be 0.25\% of $I$.

Beyond this correction it is also necessary to correct for
instrumental polarization that varies with position across the field
of view. These effects result in the conversion of $I$ into $Q$, $U$,
and $V$. In other words, sources which are unpolarized appear to be
polarized, and sources which are partially polarized appear in images
with altered polarization properties. Spurious polarized features
appear mainly wherever there is a strong point source or a strong
unpolarized extended source such as an \ion{H}{ii} region.

Two investigations have been made of this effect. The effect
has been studied by \citet{ng05} using electromagnetic analysis
software. Instrumental polarization depends on the polarization
performance of the feeds, on reflector curvature, and on aperture
blockage by the feed-support struts, among other more minor effects
such as surface roughness. Because the antennas are equatorially
mounted the instrumental polarization response remains fixed on the
field of view throughout an observation and it is possible to make an
image-based correction.

The image-based correction for instrumental polarization was derived
from observations of an unpolarized source at a grid of positions
across the main beam. Details, and images of the instrument response,
may be found in \citet{tayl03}.  The correction was a {\it{scalar}}
correction: at any position in the field, spurious $Q$ and $U$ were
calculated as a fraction of $I$ at that point, and the correction
function was determined for the aggregate of the seven
antennas. Spurious $Q$ and $U$ so derived were simply subtracted from
measured $Q$ and $U$. Individual corrections for the four frequency
bands were determined and were applied separately.

Subsequently, a more sophisticated correction technique was developed
by \citet{reid08} and used for 60\% of the survey data. The
reflector-induced instrumental polarization is not purely a scalar
quantity: it has phase as well as amplitude. Furthermore, it is quite
strongly dependent on the details of antenna construction; the
calculations of \citet{ng05} show this quite clearly. The ultimate
correction must take into account the complex nature of this response,
and to enable us to do so we made a further set of observations of the
effect. Once again, an unpolarized source was observed at a grid of
positions across the antenna main beam. These measurements were made
in a holographic mode: the reference source (in this case 3C~295) was
placed at the centre of the field of view of one antenna and the other
antennas were moved.  Complex visibilities were measured at 15\arcmin\
increments across the beam of each antenna.

The data resulting from these holographic measurements were used in
the image formation process. For each {\sc{CLEAN}} component obtained
from processing the $I$ image, spurious $Q$ and $U$ visibilities were
computed using the measured properties of the antennas participating
in each baseline.  The sum of these over all {\sc{CLEAN}} components
was then subtracted from $Q$ and $U$ visibilities, and $Q$ and $U$
images were computed from the modified visibilities.

At large angular distances from the pointing centre the conversions of
energy between the Stokes parameters increase and the primary beams
are less well known. Therefore, when making a mosaic of polarization
images the data for any one field were not used beyond a radius of
$75'$. In contrast, for total-intensity imaging the equivalent radius
is $90'$.

\section{Observation and processing of the Effelsberg 100-m data}
\label{sec:effelsberg}

\subsection{Observations}
\label{subsec:eff_obs}

Polarization data from the Effelsberg 100-m telescope are part of the
{\it Effelsberg Medium Latitude Survey} (EMLS: Reich et
al. 2004). Survey characteristics are given in
Table~\ref{tabl-100m}. 

Observations were made only at night to avoid spurious contributions
from the Sun seen in the telescope sidelobes.  The extended
observation time needed for accurate determination of zero level and
the contribution of ground radiation was not available, and the survey
area was divided into subfields of typical size
$10^{\circ}{\times}10^{\circ}$ with $5\degr$ as the smallest and
$16\degr$ as the longest scan length. Individual subfields were
treated as separate observations. Each subfield was observed at least
twice in orthogonal directions in Galactic co-ordinates with a
scanning speed of $4\deg$/min, giving a total integration time for
each $4\arcmin$ pixel of 2~sec.

\subsection{Data processing}
\label{subsec:eff_process}

A linear baseline was removed from each scan, setting $I$, $Q$, and
$U$ to zero at the ends; this effectively removed ground radiation
from the observations but also filtered off the largest
structure. Scans were calibrated against the noise source, and spiky
interference was excised.  Scanning effects were suppressed using the
method of unsharp masking \citep{sofu79}. Following \citet{junk87},
the means of $Q$ and $U$ for each scan were assumed to be
zero. Orthogonal scans were then reconciled using the ``plait''
algorithm \citep{emer88}. By assembling plaited maps for the entire
survey area we avoided edge effects at the boundaries between
subfields. Accurate observations of $I$, $Q$, and $U$ then depend on
data from smaller telescopes to provide the zero level and
large-structure information. For $I$ observations the Stockert data
\citep{reic82,reic86} provided the missing information. For $Q$ and
$U$ observations \citet{uyan98} made initial attempts to use the data
of \citet{brou76}. When this proved inadequate because of the sparse
sampling of that dataset, an initiative was launched to provide the
necessary data from the DRAO 26-m telescope, ultimately leading to the
survey of \citet{woll06}. That survey is used here.

The Effelsberg total-intensity data are confusion limited at a level
of 15~mK~$\rm T_{B}$ or about 7~mJy/beam area. The polarization
channels $Q$ and $U$ are less affected by confusion and have a typical
rms noise level of 8~mK~$\rm T_{B}$. Instrumental polarization was
reduced to a residual level of less than 1\% by the technique
described by \citet{uyan98}.

\section{Observation and processing of the DRAO 26-m data}
\label{sec:26m}

%\subsection{Observations}
%\label{subsec:26m_obs}

The data for the present work were drawn from the
absolutely calibrated survey of the polarized emission from the
northern sky at 1410~MHz made by \citet{woll06}. Details can be found in 
that paper, and only a few facts about the survey are repeated here.

Observations were made as drift scans, with the telescope stationary
on the meridian; observations were made only at night to avoid the
effects of the Sun in the telescope sidelobes. Data were fully sampled
along the scans, but the separation between scans was generally larger
than required for full sampling with a 36\arcmin\ beam. Averaged over
the survey region (spanning declination $-$29$^{\circ}$ to
+90$^{\circ}$), coverage was 42\% of full Nyquist (half-beamwidth)
sampling.

The telescope accepts both circular polarizations, and Stokes
parameters $Q$ and $U$ were derived from the complex product of LHCP
and RHCP generated in an analog correlator. Measurements of total
intensity were made from detectors in a separate signal path: $I$ and
PI therefore required separate calibration.

The standard polarization calibrator (3C~286) was used to calibrate
the Effelsberg observations, but could not be used for the 26-m
survey. Its polarized flux density, of the order of 1.5~Jy at 1.4~GHz,
generates only $\sim$140~mK in the 36\arcmin\, beam, small compared to
the polarized signal from the Galactic emission on which the source is
superimposed, and comparable to variations in that background.  Tying
the survey to the work of \citet{brou76} at 946 points across the sky
provided the required calibration of the polarization products.

The final rms noise level of the 26-m data was 12~mK in $Q$ and
$U$. Systematic errors were estimated to be less than 50~mK.

%\subsection{Data processing}
%\label{subsec:26m_process}

\section{Intercomparison of calibration of the three surveys}
\label{sec:intercomp}

Three quite disparate datasets must be combined to form the final
images and it is important to know that the three calibrations are
consistent.  The calibration of the DRAO 26-m data \citep{woll06} is
tied to the data of \citet{brou76}. The calibration of the Effelsberg
100-m data is based on the assumed polarized flux density of 3C~286
and a conversion of flux density to brightness temperature
\citep{uyan98}. The calibration of the DRAO Synthesis Telescope is
based on assumed flux densities for 3C~147 and 3C~295, both
unpolarized sources, for PI, and on 3C~286 for PA (see Section
\ref{subsec:st_cal}). In this Section we investigate the consistency
of these three calibrations.

\subsection{Comparison of the 26-m and 100-m scales}
\label{subsec:26m_100m_comp}

\begin{figure*}
\begin{minipage}{9cm}
\resizebox*{8.5cm}{!}{\includegraphics[angle=0]{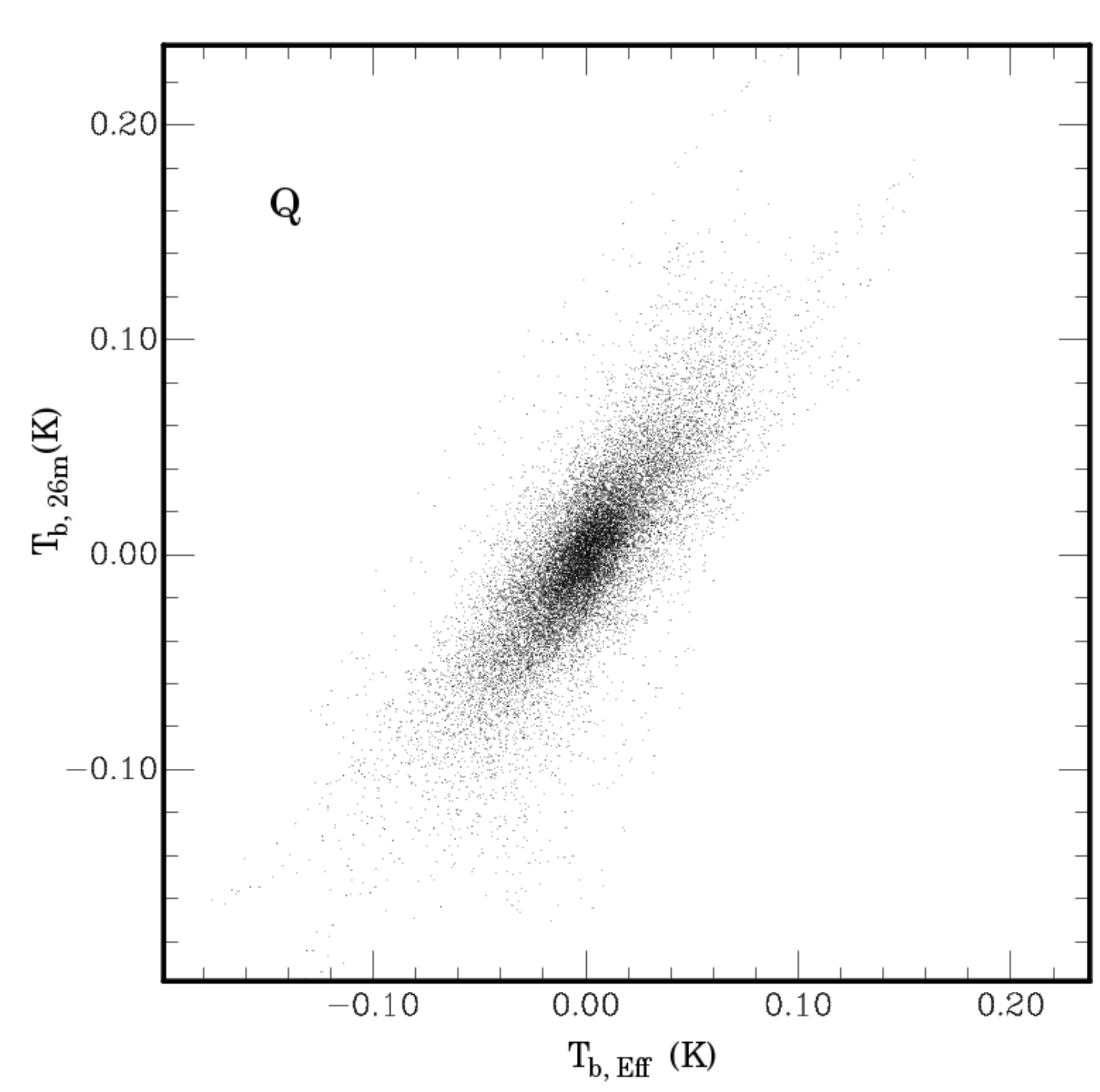}}
\end{minipage}
\hfill
\begin{minipage}{9cm}
\resizebox*{8.5cm}{!}{\includegraphics[angle=0]{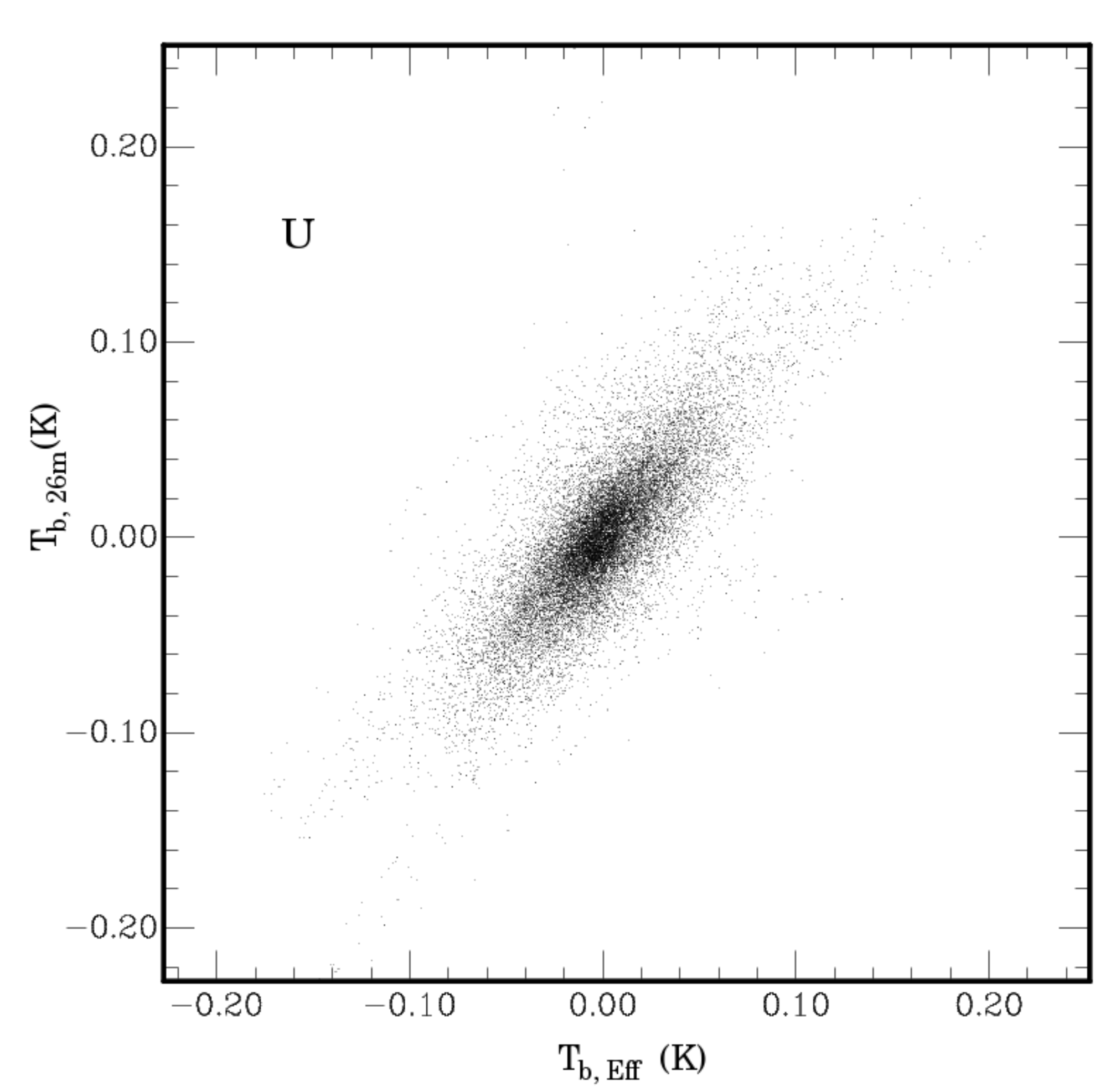}}
\end{minipage}
   \caption{Point-by-point comparison of brightness temperature values
            in images derived from Effelsberg 100-m data (x-axis) and
            DRAO 26-m data (y-axis) over an area of 726 square
            degrees. Axis scales are linear. Data have been filtered
            to include only structure of size between $36'$ and
            $2^{\circ}$. $Q$ data are shown in the left panel and $U$
            data in the right.  See text for details.}

   \label{m26_vs_eff}
\end{figure*}

The calibration of the DRAO 26-m survey is tied to the Dwingeloo survey, itself calibrated using an assumed flux density for Cas~A and a measured value of antenna gain \citep{brou76}. \citet{woll06} compared the DRAO 26-m and Effelsberg scales over two regions 1200 and 640 square degrees in area. The Effelsberg data were
smoothed to 36\arcmin, the resolution of the 26-m Telescope.  Scans
across the smoothed data in right ascension were derived, and compared
with actual scans made with the DRAO 26-m Telescope.  The two sets of
scans were not directly comparable because of the suppression of
large-scale features in the Effelsberg data, but it was assumed that the
missing data could be approximated by polynomials.  The coefficients
of these polynomials, of orders 1 to 3, and a scaling factor for the
26-m data were derived by a fitting procedure. The derived scaling
factor was 0.94, indicating that the scale of \citet{brou76} was 6\%
too high. The origin of
this discrepancy is not clear, but it appears to be within the errors of the Dwingeloo survey, especially taking into account subsequent refinements of the flux density scale 
\citep{baar77,ott94}. 

We have compared the two datasets in a different way, using data from
a region of area 726 square degrees. Images from both datasets
were prepared including only structure between 36\arcmin and
2$^{\circ}$ in size. Great care was taken to eliminate edge effects by
starting with data from larger regions before any spatial filtering
was attempted. $Q$ and $U$ images were processed independently.
Filtered images from the two datasets closely resemble each other.  A
pixel-by pixel comparison is shown in Fig.~\ref{m26_vs_eff}.
Regression lines fitted to these data do not have a slope of exactly
1.0 and do not pass exactly through the point (0,0). We cannot explain
this small effect and have therefore not used regression analysis, but
have simply taken the ratio of all the data points shown. This ratio
is $0.96{\pm}0.01$ (26-m/100-m).  We conclude that the two intensity
scales are within $\sim$5\%.

\subsection{Comparison of the 100-m and Synthesis Telescope scales}
\label{subsec:100m_st_comp}

\begin{figure*}
\begin{minipage}{9cm}
\resizebox*{9cm}{!}{\includegraphics[angle=0]{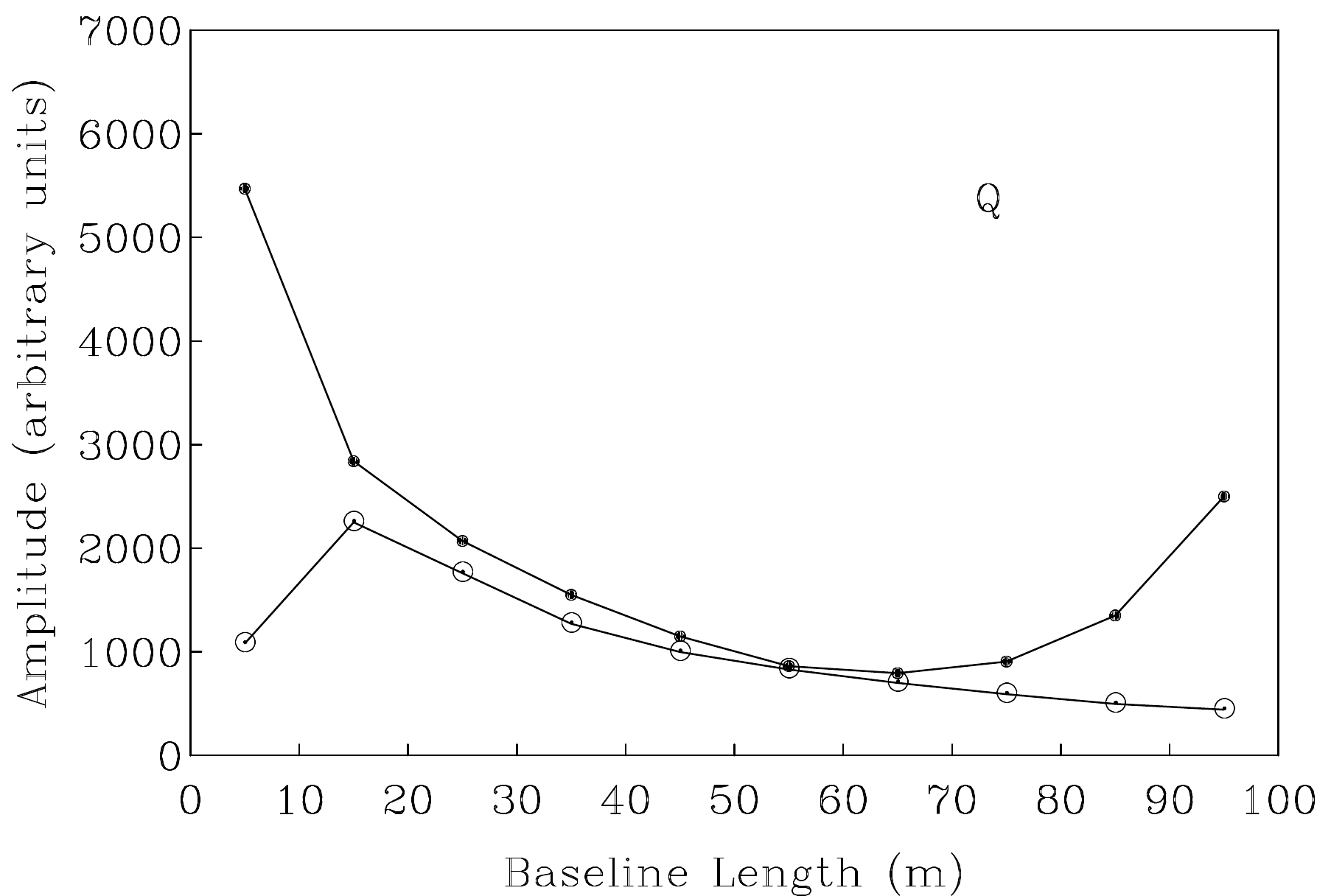}}
\end{minipage}
\hfill
\begin{minipage}{9cm}
\resizebox*{9cm}{!}{\includegraphics[angle=0]{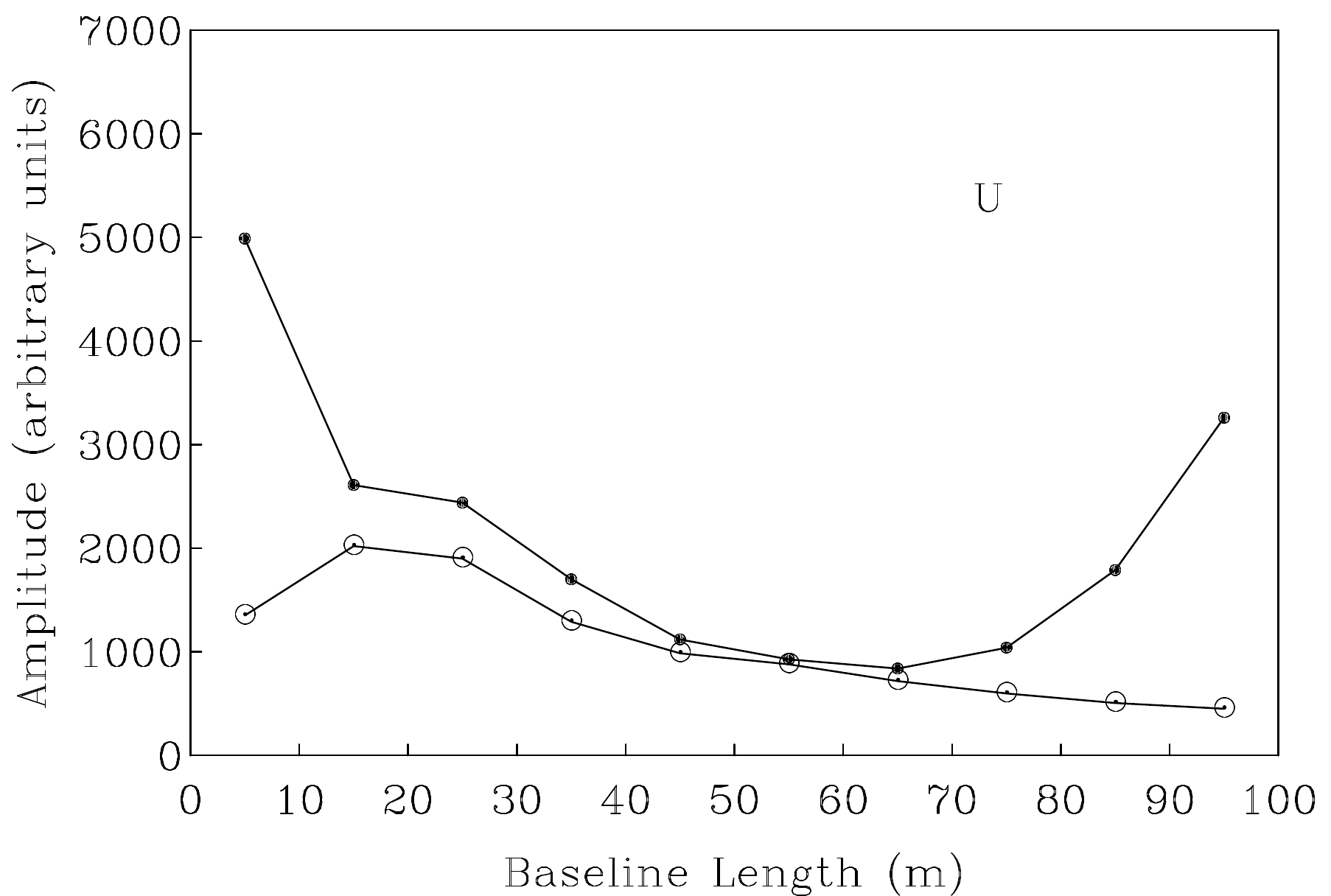}}
\end{minipage}
   \caption{Power spectra of polarization data in a region of size
            $4^{\circ} \times 4^{\circ}$ centred at
            ${\ell}={98^{\circ}}$, ${b}={1^{\circ}}$ for $Q$ (left panel)
            and $U$ (right panel) derived from data obtained with the 
            Effelsberg 100-m Telescope (filled circles) and the DRAO 
            Synthesis Telescope (open circles).  
            See text for details.}
   \label{st_vs_eff} 
\end{figure*}

The PI and PA scales of the Effelsberg 100-m Telescope and the DRAO
Synthesis Telescope are likely to be very close. The 100-m Telescope
is calibrated using 3C~286 for total intensity and PI. The Synthesis
Telescope is calibrated using 3C~147 and 3C~295. Both telescopes rely
on the same authority for the flux densities of these sources, the
work of \citet{baar77} and \citet{ott94}. Furthermore, the 15-year
experience of combining Effelsberg and DRAO data during the entire
time span of the CGPS indicates that the alignment between the
intensity scales is within a few percent.

Nevertheless, a comparison of PI and PA scales was undertaken to allay
any concern about scale errors and to check our analysis methods.  The
two telescopes have a generous overlap in the {\it{u-v}} plane, but
polarized regions are not strong emitters, so the comparison was
limited by sensitivity. The region chosen for the comparison was an
area $4^{\circ} \times 4^{\circ}$ centred on ${\ell}={98^{\circ}}$,
${b}={1^{\circ}}$ observed by both telescopes. This region contains
bright polarized emission displaying structure on both large and small
scales, and the images contain significant spectral power at the
spatial frequencies where both telescopes have significant sampling.

The data in the $4^{\circ} \times 4^{\circ}$ region were placed in a
larger image and tapered smoothly to zero at the edges. The larger
image was Fourier transformed to the visibility domain. The extension
avoided the artefacts that are introduced when transforming data with
abrupt discontinuities; the exact nature of the tapering function (a
cubic polynomial was used) did not have a strong effect on the outcome
of the comparison. The transformed 100-m data were then corrected for
the angular resolution of the telescope (9\farcm35) by dividing the
visibilities by the transform of the beam (approximated as a Gaussian
function reaching half power at 35~m, approximately one third of the
antenna diameter). The equivalent process (dividing by the transform
of the synthesized beam) was applied to the Synthesis Telescope data.
For each dataset, an approximate power spectrum was constructed by
averaging the power in the visibility plane in rings. The same
processing was applied independently to $Q$ and $U$ images.

Fig.~\ref{st_vs_eff} shows the results. At baselines below 15~m the
power in the 100-m data exceeds that from the Synthesis
Telescope. This is as expected because the interferometer cannot
sample structure on those scales. In the range 5 to 15~m the
100-m data provides the better representation of the emission
structure. At the very shortest baselines, not shown in
Fig.~\ref{st_vs_eff}, the power spectrum of the Effelsberg data drops
again, a consequence of setting values to zero at the edge of each
observation subfield.  At high spatial frequencies the corrected 100-m
power spectrum exhibits a strong rise, attributable to the
amplification of noise and high-frequency scanning artefacts that results from
dividing the visibilities by a function whose value is rapidly approaching
zero.  A single antenna obtains a noisy estimate of the image power at
its resolution limit: the finest details in an image made with a
single antenna are measured with very low weight.

Inspection of Fig.~\ref{st_vs_eff} shows that the two datasets are in
good agreement over the baseline range from about 15~m to 70~m. The
ratio (100-m to Synthesis Telescope) is very close to unity at 55~m
and averages ${1.15} \pm {0.2}$ in the range 15 to 50~m where both
datasets are used. The uncertainty was estimated from careful
consideration of the main source of error, our limited knowledge of
the beamshapes. As noted above, both sets of visibilities have been
divided by the transform of the beam. For the Effelsberg data this
process boosts visibilities at 35~m by a factor of 2 and by larger
factors at larger baselines. The Effelsberg beam has been approximated
by a Gaussian which matches the observed beam well within the
half-power beamwidth, but does not fit the lower levels well and has a
different solid angle than the actual beam. The Synthesis Telescope
beam varies slightly from place to place because of data loss and
other factors, but the effect on the comparison is smaller.
Considering these facts, and the noise level of the data, as well as
our initial confidence in the calibration reliability, we made no
further scale corrections in combining the data from the 100-m
Telescope and the Synthesis Telescope. There is no systematic
imbalance between the $Q$ and $U$ ratios, implying that the PA scales
for the two telescopes agree. We conclude that the scales of the two
datasets are probably within 10\%.

\section{Combining the three datasets}
\label{sec:combining}

Data from all three telescopes are needed to create a high-fidelity,
high-resolution image of the polarized sky.  The 26-m Telescope
provides measurements with correct zero levels, but has poor
resolution. The Effelsberg data have better resolution but lack
information on structures larger than a few degrees because of the
manner in which data were taken and ground radiation removed. The
Synthesis Telescope has good angular resolution but information
corresponding to broad structures cannot be sampled by the
interferometer. The combination process applies the appropriate
spatial filtering to account for the transfer functions of the
individual telescopes.

Fourier-transform techniques are regularly used to combine
single-antenna and aperture-synthesis data in the CGPS (see Taylor et
al. 2003 for details).  However, the 26-m polarization dataset
presents the problem of undersampling. One choice is to use the
interpolated image (as described by Wolleben et al. 2006) but after
combination with the Effelsberg data artefacts remained that were
obviously related to the undersampling. Fourier-transform techniques
were ruled out, but we set out to devise a process that exploited the
information content of each dataset to the maximum extent possible.

\begin{figure}
%   \centerline{\includegraphics[bb = 70 100 570 450,width=8.5cm,clip]
%    {f3.eps}}     
\resizebox*{8.5cm}{!}{\includegraphics[angle=0]{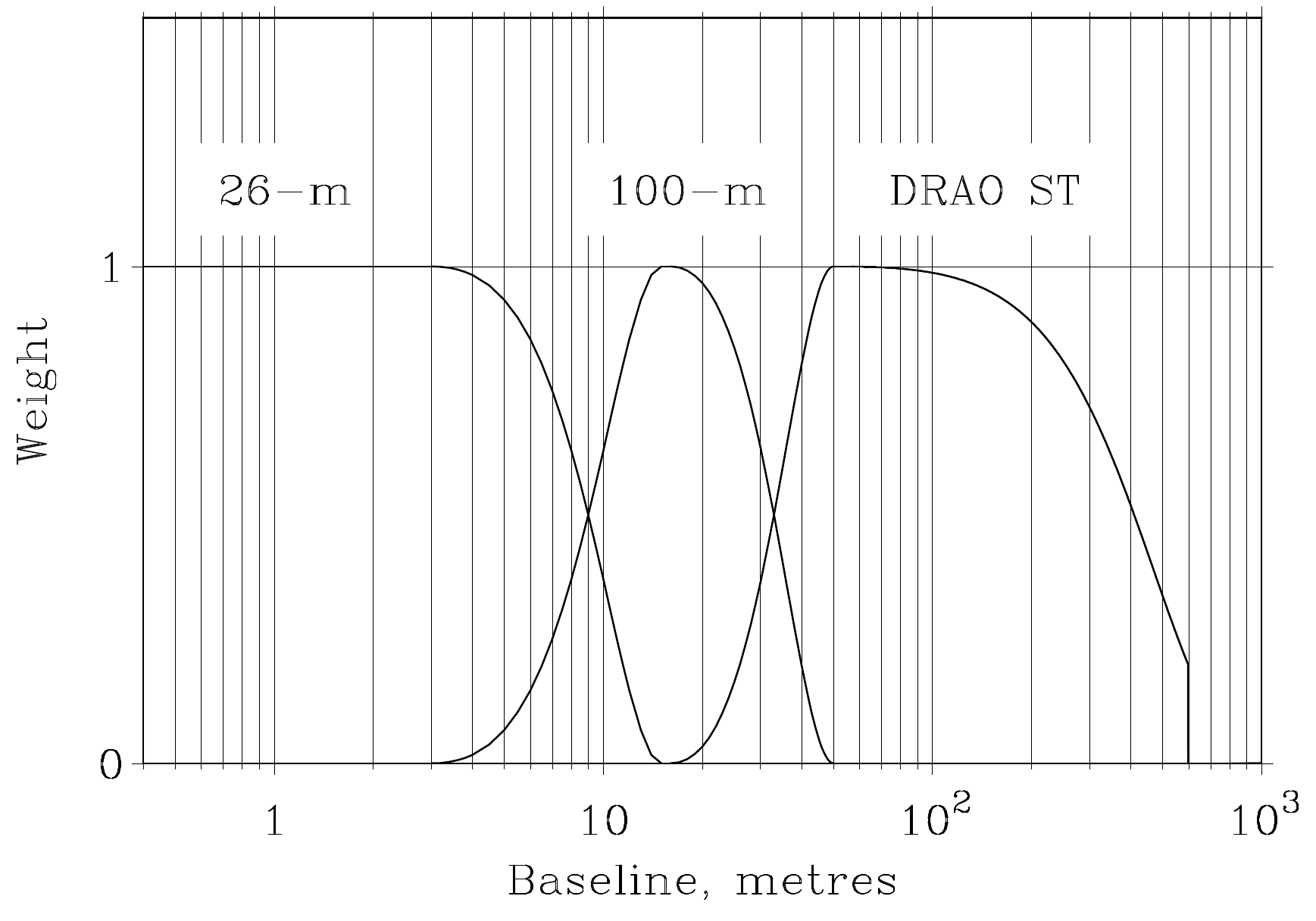}}
   \caption{Approximate spatial filtering applied in the combination
    of datasets from the DRAO 26-m Telescope, the Effelsberg 100-m
    Telescope, and the DRAO Synthesis Telescope. Because image-domain
    methods were used to combine the 26-m and 100-m datasets (see
    text) the filter functions that overlap at 9 m, although
    approximately correct, are notional.}
   \label{filter}
\end{figure}

The first step was to improve the interpolation across the gaps in the
26-m data. In place of linear or other mathematical interpolation, we
used information from the Effelsberg 100-m data. We smoothed the
Effelsberg data to $36'$. We worked in Galactic co-ordinates,
processing data along lines of constant longitude, one longitude at a
time. Where 26-m data points were available we used them. Across a gap
where no 26-m data were available, we used smoothed Effelsberg data,
``lifting'' the data to attach it smoothly to the 26-m data points at
each end. This process did not fill the gaps perfectly, but
obviously brought in more information about the sky than a simple linear
interpolation. The result of this step (which we refer to as the final 26-m
image) was our best estimate of the sky seen with a 26-m telescope. We
then had a dataset which sampled structure with a filtering function
whose half-power level was at about 9 metres, or 45 wavelengths (see
Fig.~\ref{filter}).  This dataset was then combined with the
Effelsberg 100-m dataset using a process described by
\citet{reic90}. The Effelsberg data were smoothed to the resolution of
the final 26-m image, and the difference computed. The difference
represents structure that is not present in the original Effelsberg
image (the broad structure filtered off by the observing technique),
and this was then added to the Effelsberg data. The result of this
step is to restore the broad structure that is removed in the Effeleberg
data reduction procedure.

Fig.~\ref{filter} shows the weighting functions used in the
combination in a diagrammatic fashion.  The process that we used to
combine 26-m and Effelsberg data on average blended the data between
baselines 3 m and 15 m (3 m, about 15 wavelengths, corresponds to an
angular size of 4$^{\circ}$, the smallest area mapped in the
Effelsberg survey). The ideal weighting function is at 0.5 at 9~m,
matching closely the weighting of the 26-m beam.  Fourier transform
methods were used to combine the [26-m $+$ 100-m] dataset with the
Synthesis Telecope data over the range of 15 to 50 m. The weighting
function for single-antenna data is at 0.5 at $\sim$33~m, again very
close to the weighting of the 100-m beam. Also shown in
Fig.~\ref{filter} is the Gaussian apodization function, reaching 20\%
at the longest baseline (617~m), that has been applied to the
Synthesis Telescope data.

\subsection{Spurious Features and Error estimates}
\label{subsec:error_est}

Some spurious features remain in the images, originating in all three
datasets. Some image defects in the Synthesis Telescope data
(ring structures) can be seen around the position of Cas~A
(G111.7$-$2.1). Similar rings can be seen around the position of the
\ion{H}{ii} region W3 (G133.7$-$1.2). Cas~A could not be imaged with
adequate dynamic range by the Synthesis Telescope and an area around
it was set to zero in that dataset. Sidelobe responses are also very
evident in the Effelsberg data around Cas~A.

Residual instrumental polarization in the Effelsberg data is less than
1\% \citep{reic04}.  Residual errors in instrumental polarization for
an individual field from the Synthesis Telescope are estimated at
0.25\% for the field centre, growing to 1\% at a distance of
$75'$. Instrumental polarization over the vast majority of the area
presented here is below 0.5\%. These estimates are based both on the
expected accuracy of the measurements for the correction, and the
observed discrepancies between overlapping fields.  Noise on Synthesis
Telescope data was 0.30~mJy/beam rms in mosaicked images for the
earlier part of the survey and 0.23~mJy/beam in the later stages,
corresponding to 76 and
58~$\thinspace{\rm{sin{\thinspace}(declination)}}$~mK respectively
(see Table~\ref{tabl-st}).

Our investigation (Section~\ref{sec:intercomp}) of the relative
calibration of the three datasets has led to the conclusion that (a)
the discrepancy between the 26-m and 100-m scales is less than 5\%, and
(b) the discrepancy between the 100-m scale and the aperture-synthesis
scale is less than 10\%. An unknown error remains in the
single-antenna data as a result of the undersampling present in the
26-m Telescope data.

The noise on the 26-m Telescope data is 12~mK rms and systematic
errors are $\le$50~mK \citep{woll06}. These errors transfer to the
final data presented here. Polarized intensity images presented in
this paper have not been corrected for noise bias.

The three datasets were observed at slightly different centre
frequencies and with different bandwidths, possibly leading to
incorrect results. We considered four sources of error, differing
bandwidth depolarization, systematic differences in polarization angle
at different frequencies, intrinsic variations in polarized intensity
with frequency, and variations in polarization angle of calibration
sources.

For a value of
${{\vert}{\rm{RM}}{\vert}}=100\thinspace{\rm{rad\thinspace{m}}}^{-2}$,
the highest value observed in a Rotation Measure synthesis study of
the Northern sky by \citet{woll10}, the bandwidth depolarization in
the four bands of the Synthesis Telescope averaged together is about
0.15\%. Bandwidths used with the Effelsberg 100-m and DRAO 26-m
Telscopes are smaller, and bandwidth depolarization is correspondingly
less. For the same values of RM, the angle rotation between the
different centre frequencies has a maximum value of
$\sim$7.5$^{\circ}$. The RM of the great bulk of emission is far less
than 100{\thinspace}rad{\thinspace}{m}$^{-2}$, probably less than
20{\thinspace}rad{\thinspace}{m}$^{-2}$, making these errors
negligible in comparison to other effects.

Assuming a temperature spectral index of 2.7 for synchrotron emission
at our frequency, the 20~MHz spread in centre frequencies makes a
difference of $\sim$4\% between amplitude scales. This may explain
some of the ratio of $\sim$1.15 found between Effelsberg 100-m and
DRAO Synthesis Telescope scales (Section \ref{subsec:100m_st_comp})
but the uncertainties are high and we have not made any adjustment to
the intensity scales. The flux densities used for the different
amplitude calibrators were appropriate for the operating frequencies
of the individual telescopes and no errors arise from the difference
in observing frequencies beyond the effect of spectral index variation
of the emission itself. The primary calibrator of polarization angle,
3C286, has almost zero RM, making its polarization angle nearly
independent of frequency.

\section{Results}
\label{sec:results}

This section presents the results of the survey with some preliminary
interpretation.  The data presented in this paper, together with data 
for other significant ISM constituents, are available at
{\tt{http://www3.cadc-ccda.hia-iha.nrc-cnrc.gc.ca/cgps/}}. Details of
the data format may be found in \citet{tayl03}.

\subsection{The images}
\label{subsec:images}

\begin{figure*}
%\begin{minipage}{4.9cm}
\begin{minipage}{4.9cm}
%  \centerline{\includegraphics[bb = 70 12 265 805,width=4.9cm,clip]{f4a.eps}}
\resizebox*{5cm}{!}{\includegraphics[angle=0]{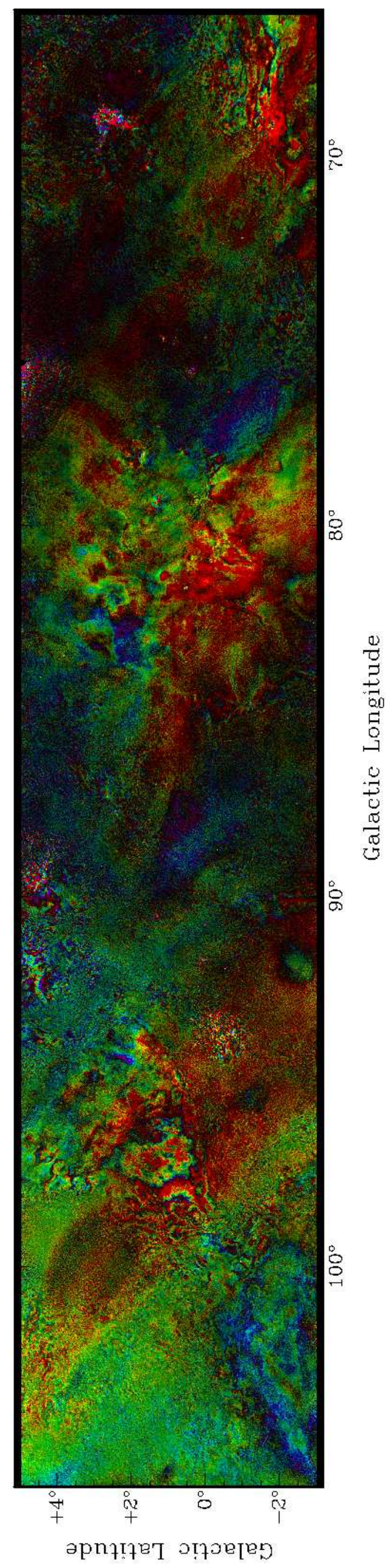}}
\end{minipage}
\hfill
\begin{minipage}{4.9cm}
%  \centerline{\includegraphics[bb = 70 12 265 805,width=4.9cm,clip]{f4b.eps}}
\resizebox*{5cm}{!}{\includegraphics[angle=0]{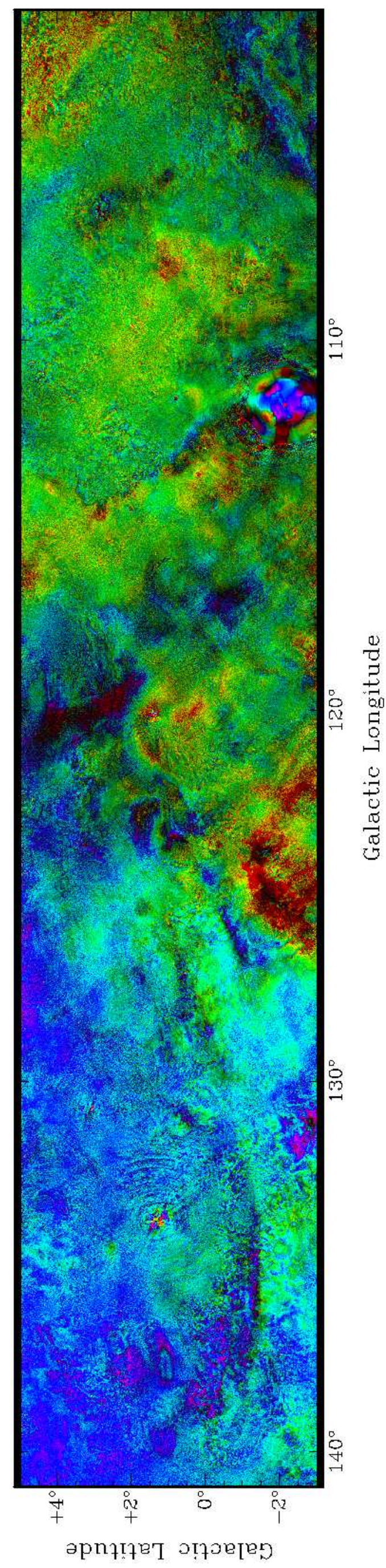}}
\end{minipage}
\hfill
\begin{minipage}{4.9cm}
%  \centerline{\includegraphics[bb = 70 12 265 805,width=4.9cm,clip]{f4c.eps}}
\resizebox*{5cm}{!}{\includegraphics[angle=0]{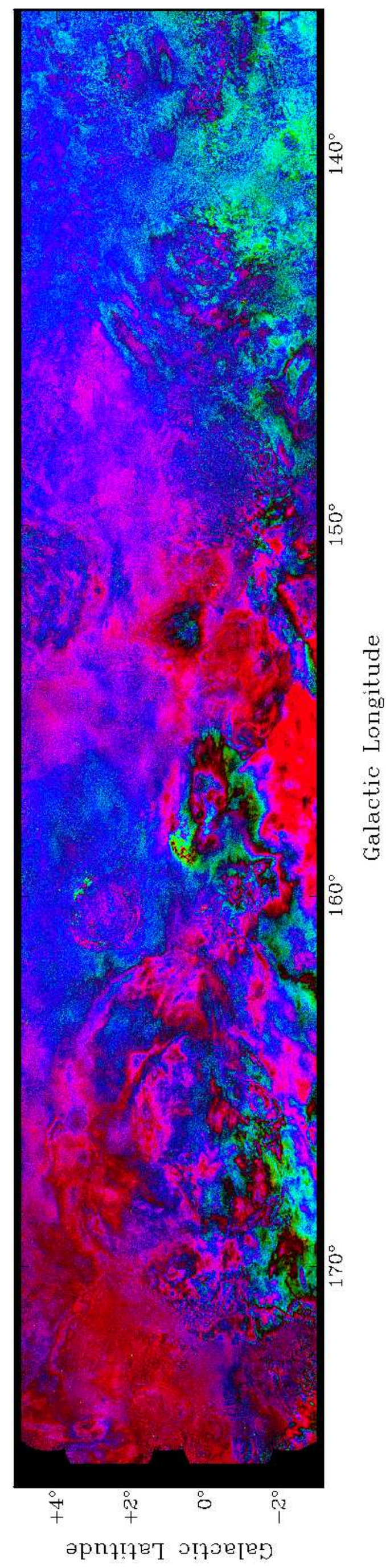}}
\end{minipage}
\caption{Polarization data along the Galactic Plane. In this
representation intensity of colour depicts polarized intensity and hue
depicts polarization angle. The range of polarized intensity is zero
(dark) to 0.45~K (bright). The range of polarization angle is from
$-50^{\circ}$ to $+30^{\circ}$, chosen because most values of angle in the
region lie in this range. Regions near ${\ell}={65^{\circ}}$ have
${\rm{PA}}{\approx}{-50^{\circ}}$ and are shown in red. Angle changes
smoothly through orange, yellow, green, blue, and purple, to red again
near ${\ell}={175^{\circ}}$ where ${\rm{PA}}{\approx}{+30^{\circ}}$.}
   \label{whole_survey}
\end{figure*}

\begin{figure*}
%   \centerline{\includegraphics[bb = 69 119 478
%   728,width=14.5cm,clip]{f5.eps}}
\resizebox*{20cm}{!}{\centerline{\includegraphics[angle=0]{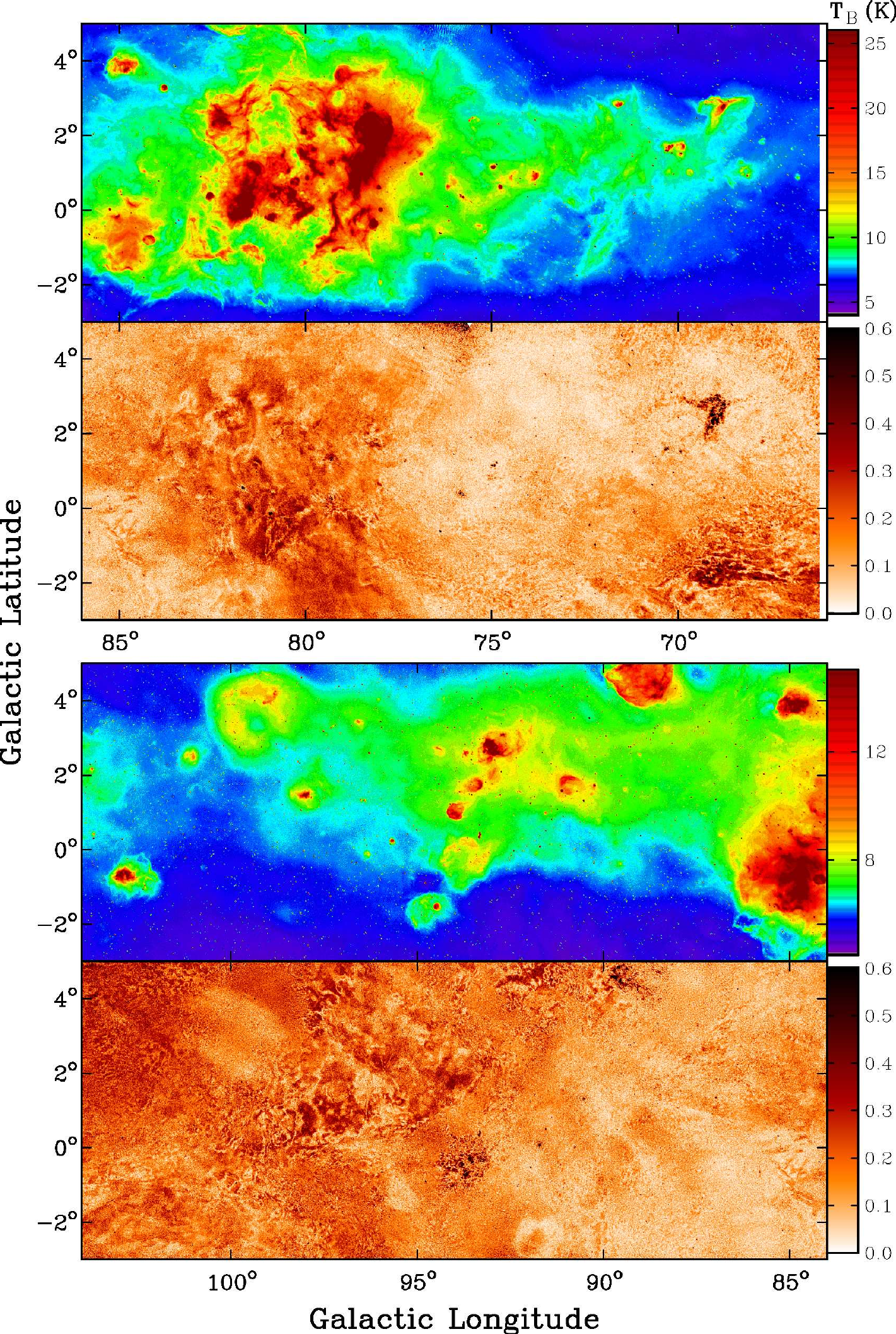}}}
   \caption{Maps of total intensity (multi-colour) and polarized 
   intensity (orange) in the regions 
   ${66^{\circ}}<{\ell}<{86^{\circ}}$ (top), and
   ${84^{\circ}}<{\ell}<{104^{\circ}}$ (bottom).}
   \label{ipi66104}
\end{figure*}

\begin{figure*}
%   \centerline{\includegraphics[bb = 69 119 478
%   728,width=14.5cm,clip]{f6.eps}}
\resizebox*{20cm}{!}{\centerline{\includegraphics[angle=0]{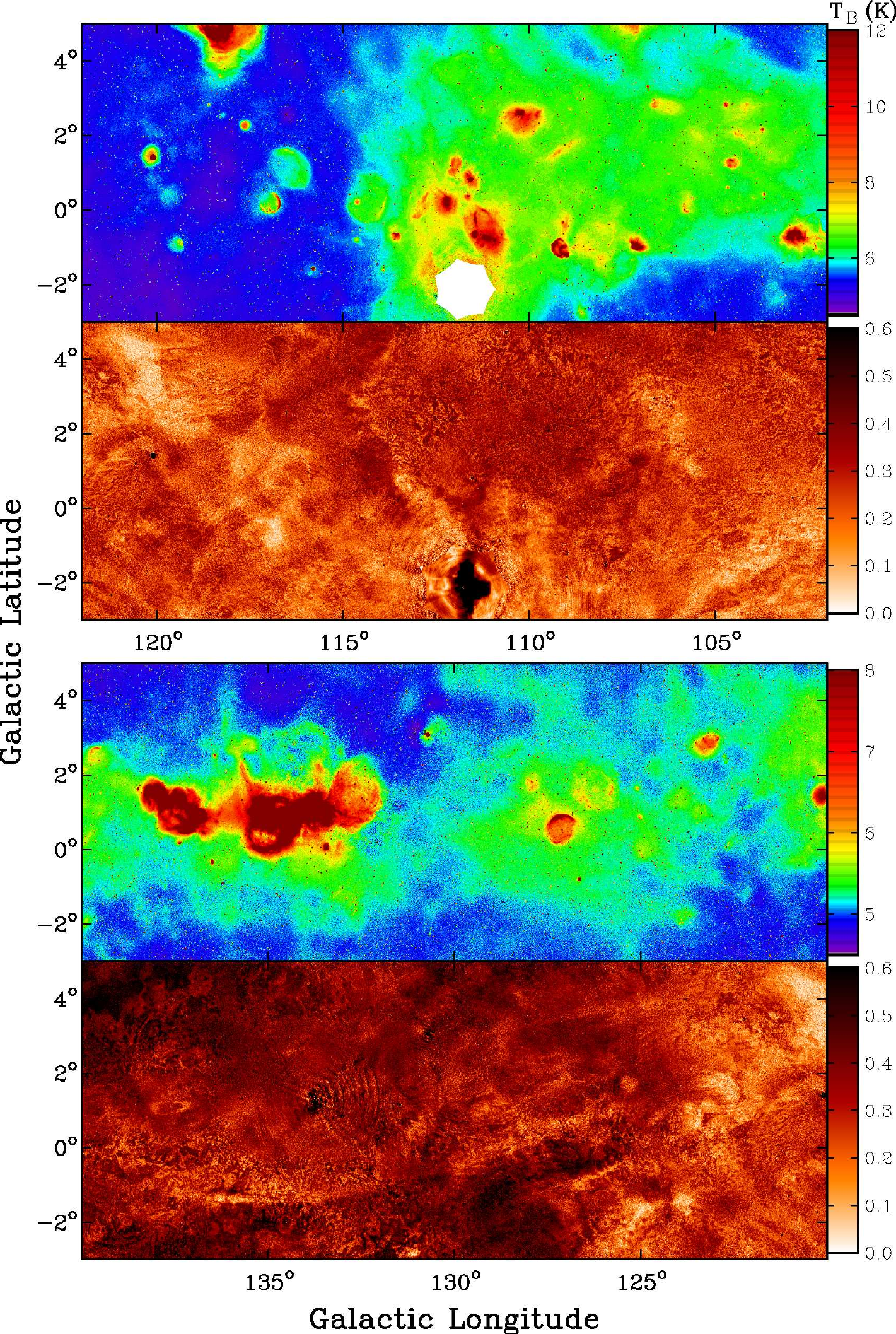}}}
   \caption{Maps of total intensity (multi-colour) and polarized 
   intensity (orange) in the regions
   ${102^{\circ}}<{\ell}<{122^{\circ}}$ (top), and
   ${120^{\circ}}<{\ell}<{140^{\circ}}$ (bottom).}
   \label{ipi102140}
\end{figure*}

\begin{figure*}
%   \centerline{\includegraphics[bb = 69 119 478
%   728,width=14.5cm,clip]{f7.eps}}
\resizebox*{20cm}{!}{\centerline{\includegraphics[angle=0]{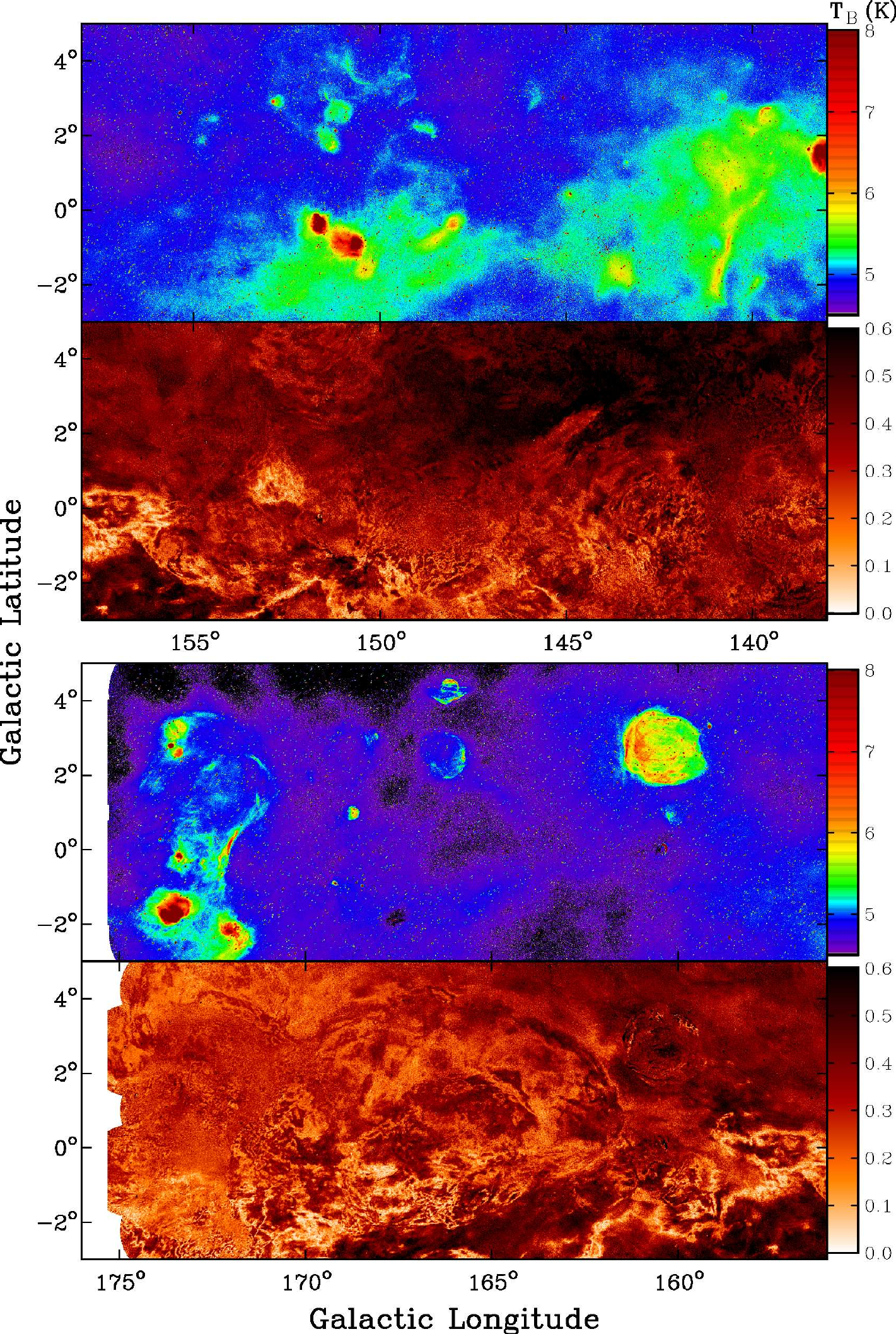}}}
   \caption{Maps of total intensity (multi-colour) and polarized 
   intensity (orange) in the regions
   ${138^{\circ}}<{\ell}<{158^{\circ}}$ (top), and
   ${156^{\circ}}<{\ell}<{176^{\circ}}$ (bottom).}
   \label{ipi158176}
\end{figure*}

Figure~\ref{whole_survey} shows the major part of the survey, along
the Galactic plane, using a representation that shows both PI and PA.
Figs.~\ref{ipi66104}, \ref{ipi102140} and \ref{ipi158176} present
$I$ and PI for the same area. Each panel covers 20$^{\circ}$ of
longitude, with an overlap of 2$^{\circ}$ between
figures. Figure~\ref{highlat} presents the data for the high-latitude
extension, again showing PI and PA. The survey data are shown at full
angular resolution.

We are presenting the largest survey yet of the polarized emission
from the Milky Way and the first extensive survey to combine
single-antenna data with aperture-synthesis data. We expect new
results from such an advance, and indeed we have discovered a number
of surprising phenomena. Here we attempt no more than a broad-brush
interpretation, emphasizing major themes that appear to be new while
paying relatively little attention to particular objects. Future
papers will deal with many different results from this survey.

Over the region covered by our data the diffuse polarized sky bears
little resemblance to the total-intensity sky. A similar remark has
been made about virtually every polarization survey undertaken at
decimetre wavelengths throughout the history of the subject. The
widely accepted interpretation is that the polarized features we
detect are largely the product of Faraday rotation and do not reflect
structure in the Galactic synchrotron emission. The exceptions are
supernova remnants (SNRs) and pulsar-wind nebulae and, in particular,
the North Polar Spur, seen at high positive latitudes,
{\it{e.g.}}~\citet{woll06,woll07} (and not covered in our survey), and
``point'' sources, mostly extragalactic. SNRs are discussed in
Section~\ref{subsec:snrs}. Point source RMs are discussed in detail by
\citet{brow01} and \citet{brow03} and will not be considered here.

An origin in Faraday rotation means automatically that polarized
``objects'' will usually not look like things seen in other
wavebands. The polarized sky is almost entirely ``new'', and the first
problem we should tackle is taxonomy, a classification of polarization
features.  We will not attempt that in any thorough way here, but our
choice of the small number of features that we do describe may point
the way for a future study.

\subsection{Large-scale features}
\label{subsec:large-scale}

\begin{figure}
%   \centerline{\includegraphics[bb = 62 83 550 421,width=8.5cm,clip]
%    {f8.eps}}     
\resizebox*{8.5cm}{!}{\includegraphics[angle=0]{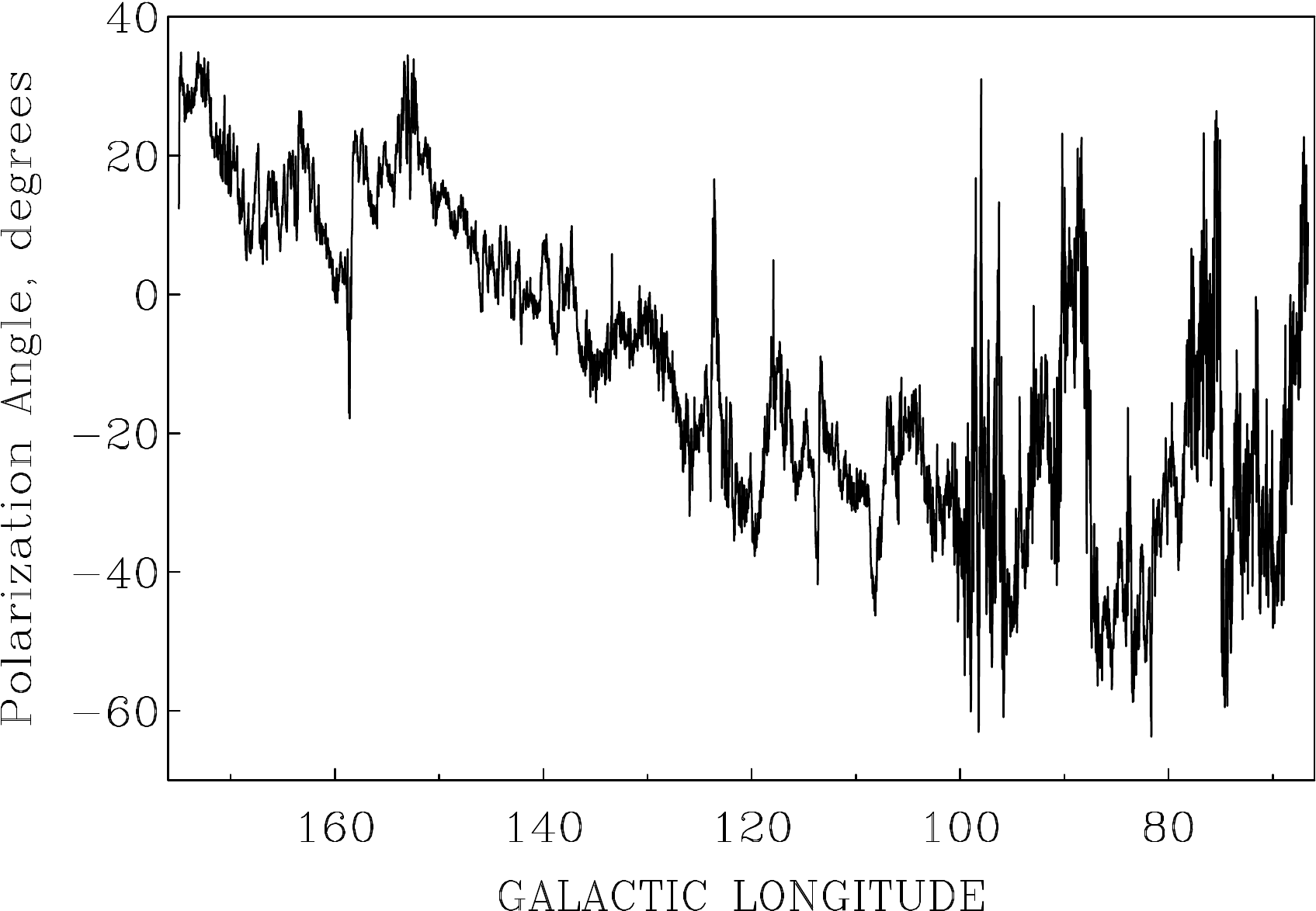}}
   \caption{A representative plot showing variation of polarization
            angle averaged over a strip $20'$ wide centred on 
            ${b}={1^{\circ}}$.}
   \label{angle}
\end{figure}

We begin with a description of some of the large-scale features of the
data seen in Fig.~\ref{whole_survey}. There is a gradual increase in
PI and a fairly smooth change of PA of about 70$^{\circ}$ with
increasing longitude. The angle variation is illustrated in
Fig.~\ref{angle}.  The Cygnus region obviously ceates a major
disturbance in the range ${65^{\circ}}<{\ell}<{100^{\circ}}$ but the
large-scale trend is nevertheless clear.  These trends were already
evident in the DRAO 26-m data \citep{woll06}, and are preserved at
high angular resolution.  The large-scale changes are probably
connected with the large-scale features of the magnetic field, roughly
aligned with the spiral structure, while the small-scale polarization
structures reflect the irregular component and the turbulence of the
ionized gas (field is probably tied to electron density to some
extent).  At our observing frequency of 1.4~GHz, the irregular
component does not seem to dominate the regular one, at least for
${\ell}{\geq}{110^{\circ}}$.  At longer wavelengths, where Faraday
rotation is larger, the small-scale fluctuations may eventually
overwhelm the large-scale variation.

Such systematic changes over a large fraction of the Galactic disk
imply that the large-scale magnetic field affects the diffuse Galactic
emission, either through the emission process or through depolarization
mechanisms. The RMs of extragalactic compact sources seen
through this part of the disk have also been shown to change in a
systematic fashion \citep{brow01,brow03}. The regular component of the
local magnetic field is directed towards ${\ell}{\approx}{85^{\circ}}$
\citep{nout09}. Along lines of sight close to this direction Faraday
rotation will be high and a number of depolarization mechanisms will
be active ({\it{e.g.}} Sokoloff et al. 1998). As the line of sight
moves towards the anticentre it gradually becomes perpendicular to the
regular component of the magnetic field and Faraday rotation will
decrease. Synchrotron emission is lower in this part of
the Galaxy as shown by lower total-intensity levels, but depolarization 
effects apparently drop off even more rapidly, with the net result that 
PI rises and polarization fraction increases. The reduced amount of 
interstellar matter towards the edge of the Galaxy will also significantly 
lower depolarization.

The angle gradient seen in Fig.~\ref{angle} is consistent in sense
with the change in RM of extragalactic sources seen in \citet{brow01}
and \citet{brow03} but the change in angle of $\sim$70$^{\circ}$
between ${\ell}\approx{100^{\circ}}$ and ${\ell}\approx{180^{\circ}}$
is less by a factor of about 8 than the rotation implied by the change
in RM over the same range, a change of
$\sim$200{\thinspace}rad{\thinspace}m$^{-2}$.  The data of
\citet{woll06} reveal strong depolarization along this section of the
Galactic plane, so we are not seeing all of the polarized
emission. The large-scale emission, which dominates the polarization
angle, is probably mostly local emission (see discussion of
depolarization in Section~\ref{subsec:small-scale}).

\subsection{Small-scale structure and depolarization}
\label{subsec:small-scale}

Chaotic structure, apparently uncorrelated with any other ISM tracer,
has been the hallmark of high-resolution polarization images since
they were first made (see the references in the Introduction to this
paper). The present survey fits this pattern. Our results
(Figs.~\ref{ipi66104}, \ref{ipi102140}, and \ref{ipi158176}) show
that small-scale structure is seen almost everywhere in our data.

Faraday rotation that varies with position can break large, smooth
emission features into smaller polarization structures. Faraday rotation
in a turbulent medium will therefore affect the power spectrum of the
polarized sky, moving power from broad structure to smaller scales. At
1420~MHz this process must be active to some extent, but large-scale
features are still detectable and indeed are dominant, as noted in
Section~\ref{subsec:large-scale}. Faraday rotation depends on
wavelength squared and there is undoubtedly some wavelength at which
the large-scale structure has virtually disappeared from polarization
data. 

As noted in the introduction, radio telescopes at decimetre
wavelengths are often more sensitive to the Faraday rotation produced
by ionized gas than to its total-intensity emission. For example,
consider an ionized region of extent 10~pc with electron density
0.5~cm$^{-3}$ in a magnetic field with line-of-sight component
2~$\mu$G. At 1420~MHz this region will produce a Faraday rotation of
20$^{\circ}$, which is easily detectable with the DRAO Synthesis
Telescope. The same \ion{H}{ii} region produces an emission measure of
2.5~cm$^{-6}${\thinspace}pc, one tenth the thermal noise in the $I$
image. For the Effelsberg 100-m Telescope the angle change will be the
same, and also easily detectable, while the signal in the
total-intensity channel will be only 1~$\sigma$ (if the \ion{H}{ii}
region fills the beam).

Ironically, the very effect that makes ionized gas so easy to detect
can be a drawback, because copious Faraday rotation produces
depolarization. Depolarization is a reduction in observed fractional
polarization that arises from vector averaging
\citep{gray99,uyan03}. The averaging may occur (a) along the line of
sight, in which case the phenomenon is known as {\it{differential
Faraday rotation}} or {\it{depth depolarization}}, (b) within the
telescope beam ({\it{Faraday dispersion}} or {\it{beam
depolarization}}) or (c) within the receiver bandpass ({\it{bandwidth
depolarization}}).

We see two kinds of depolarization features. The first kind is
characterized by a sharp reduction of PI relative to the
surroundings. This is depolarization by a relatively close \ion{H}{ii}
region. Many of these \ion{H}{ii} regions can be recognized in
total intensity or H$\alpha$ images.  The dense turbulent ionized
material, probably accompanied by tangled magnetic field, strongly
reduces the fractional polarization of emission from greater
distances, and there is little synchrotron emission on the near
side. The second kind is characterized by sharp boundaries within
which the PI is still significant, but the structure becomes much
smoother relative to the surroundings. In our data these instances are
usually produced when a Local Arm \ion{H}{ii} region depolarizes more
distant emission arising in the Perseus Arm. The relatively smooth
polarized emission is synchrotron emission generated on the near side
of the \ion{H}{ii} region, possibly affected by Faraday
rotation. These effects can be used to estimate the distance to
polarization features, based on known distances to \ion{H}{ii} regions
\citep{koth04}.

We note, however, that 4.9~GHz observations \citep{sun07,gao10} show
that some \ion{H}{ii} regions have quite different properties. At that
frequency some faint \ion{H}{ii} regions generate significant regular
Faraday rotation and must contain fairly strong regular fields. There
is sufficient random fluctuation within these regions that they are
depolarized at 1420~MHz.  Detailed comparisons will be the subject of
a future paper.

The concept of the {\it{polarization horizon}}, introduced by
\citet{uyan03}, is relevant in this discussion. The combined effects of depth
depolarization and beam depolarization do not allow us to detect
polarized emission beyond a certain distance. The distance to the
polarization horizon depends on frequency, beamwidth, and
direction. \citet{koth04} show that, in the Cygnus direction, the
distance to the polarization horizon is about 2 kpc: we are seeing
predominantly local features between ${\ell}{\approx}{66^{\circ}}$ and
${\ell}{\approx}{100^{\circ}}$. At greater longitudes the images are a
blend of local polarization features with those in the Perseus Arm. In
general the local features are large and smooth, while the Perseus Arm
features are characterized by small-scale structure that appears
entirely random. Towards the anticentre the polarization horizon is well
beyond the Perseus Arm, possibly even beyond the outer limits of the Galaxy.

\subsection{Supernova remnants}
\label{subsec:snrs}

Some SNRs appear as prominent polarized objects in the survey. The
polarized signatures of CTB80 (G69.0+2.7), HB21 (G89.0+4.7), CTB104A
(G93.7$-$0.2), Tycho's SNR (G120.1+0.4), 3C58 (G130.7+3.1), and HB9
(G160.9+2.6) are very clear. The emission from the direction of many
SNRs is polarized but the polarized signal may not necessarily be SNR
emission. First, careful subtraction of the local diffuse Galactic
polarization is required to isolate any polarized emission from the
SNR itself. Second, a polarized background may suffer Faraday rotation
within the SNR.  \citet{koth06} have made a thorough study of the 34
SNRs detected in the area of the data presented in this paper; they
detected significant polarized emission from 18 of them. If the SNR is
beyond the polarization horizon its polarized emission can not usually
be detected \citep{uyan03}.

\subsection{The Cygnus-X region}
\label{subsec:cygx}

The Cygnus-X region, whose radio emission peaks at
${\ell}={80^{\circ}},~{b}={1^{\circ}}$, is the Local spiral arm seen
end-on: in this direction many objects at distances 1 to 4~kpc lie
close together on the sky \citep{wend91}.  Cygnus X is a strong
emitter in total intensity, with an integrated flux density of about
1000~Jy at 1.4~GHz from a region roughly ${6^{\circ}} \times
{6^{\circ}}$ in size, and its emission is dominated by thermal
emission \citep{wend91,knod00} although a few non-thermal sources are
also present. Surprisingly, there is significant polarized emission
roughly coincident with the total-intensity emission from Cygnus X,
and the PI along the Galactic plane on either side of the emission
complex (from ${{\ell}{\approx}70^{\circ}}$ to
${{\ell}{\approx}77^{\circ}}$ and from ${{\ell}{\approx}84^{\circ}}$
to ${{\ell}{\approx}91^{\circ}}$) is much lower (see
Fig.~\ref{whole_survey} and Fig.~\ref{ipi66104}).  The same effect can
be seen in all three datasets, the DRAO 26-m data, the Effelsberg
data, and the DRAO Synthesis Telescope data: it occurs on all size
scales.

At first sight it is puzzling that a thermal region can apparently
produce polarized emission, but this can be interpreted by
understanding the depolarization process. The ionized gas in Cygnus~X
is dense and turbulent, and, since we expect the frozen-in field to
share these characteristics, Cygnus~X will effectively scramble the
polarization of any synchrotron emission that is generated behind
it. The objects that comprise Cygnus X are concentrated between
distances of 1 and 4~kpc. The polarized emission that is seen in
Fig.~\ref{ipi66104} towards Cygnus~X must therefore be generated and
Faraday rotated along the nearest 1~kpc of the line of sight. On the
other hand, along lines of sight to either side of Cygnus~X there is
no large concentration of depolarizing material, and emission
generated along the entire line of sight is superimposed, and vector
averaging over long paths leaves little polarized signal.  This
general direction is nearly aligned with the direction of the local
magnetic field and we expect strong depth depolarization here.

The effect described can be exploited to study the local
synchrotron emissivity (on this and on other lines-of-sight that
intersect dense \ion{H}{ii} material). This is a topic beyond the
scope of the present paper, but we note that a study over a range of
wavelengths would be productive.

\subsection{The HB3/W3/W4/W5 complex}
\label{subsec:w543}

The HB3/W3/W4/W5 complex (${\ell}{\approx}{132^{\circ}}$ to
${\ell}{\approx}{139^{\circ}}$ at ${b}{\approx}{1^{\circ}}$) lies on
the near side of the Perseus Arm at a distance of $\sim$2.2~kpc
\citep{norm96}. Polarization features in the vicinity are well
studied.  Small-scale disorganized structure is depolarized by
the \ion{H}{ii} regions W3, W4, and W5 \citep{gray99} implying that
it is generated by Faraday rotation in the Perseus Arm.  

\citet{gray98} discuss a lens-shaped feature seen in
Fig.~\ref{ipi102140} superimposed on W5 (${\ell}={137\fdg6},
{b}={1\fdg1}$) and place it in the interarm at a distance of
approximately 1~kpc. It is remarkable for its highly ordered structure
in an area where all other polarization structure is disordered.  From
the Synthesis Telescope alone \citet{gray98} found $\Delta{\rm{PA}} =
{280^{\circ}}$ from edge to centre of this object requiring a change
of RM of ${\sim}$110\,rad\,m$^{-2}$. From the present data
we find $\Delta{\rm{PA}} = {\sim}25^{\circ}$, implying a much
smaller $\Delta$RM.  This illustrates the problem inherent in
interpreting aperture-synthesis data without the addition of
single-antenna data. The general conclusions reached by \citet{gray99}
about this object are unchanged by the smaller $\Delta{\rm{PA}}$ from
our new data but the distance constraints are somewhat loosened so
that the object could be almost anywhere along the 2.2~kpc line of
sight to W5.

With the wide spatial coverage of the images presented here we see a
new ``object'' that may be related to HB3/W3/W4/W5, a large curved
polarization feature that lies just south of the complex
(Fig.~\ref{ipi102140}). It is detectable from
${\ell}{\approx}{132^{\circ}}$ to ${\ell}{\approx}{140^{\circ}}$ at
${b}\approx{-1^{\circ}}$. There is no doubt that this arc-like feature
is real: it is detectable in the three individual data sets that have
gone into making the survey images. Part of the arc is seen in the
data presented by \citet{gray99}, who commented that it might be
related to W4.  We refer to this feature as the W4 Polarization
Arc. It shows many characteristics of a shock front. First, it is
curved, and its curvature can be fitted approximately by a circle of
radius $\sim$9$^{\circ}$. Second, it has a steep outer edge (to larger
radius) and a relatively shallow inner edge. If it is a shock, its
polarization signature is the result of additional Faraday rotation in
compressed material.

We have no information on its distance. It is either a feature of the
local ISM or it is in the Perseus Arm. No association has yet been
found with other ISM tracers that might allow us to discriminate
between these two choices. For the sake of discussion, we will assume
nominal distances for the two possibilities of 100~pc and 2.2~kpc, the
latter value the distance to the W3/W4/W5 \ion{H}{ii} regions
\citep{norm96}.

The large angular extent and smooth structure of the Arc may favour a
small distance. At a distance of 100~pc the length of the Arc would be
about 25~pc. Shock fronts ({\it{e.g.}}\,SNRs) do not generally reach
sizes much larger than this without severe distortion by ISM
irregularities. If there is a complete spherical shock front its
diameter would be about 30~pc. We would expect a local SNR of this
size to have left detectable traces; their absence leads us to
discount this interpretation.

If the Polarization Arc is a Perseus Arm feature at a distance of
2.2~kpc its size is ${\sim}$550~pc. While this is very large, there is
evidence that may associate it with the W3/W4/W5 \ion{H}{ii}
complex. As explained above, the small-scale polarization
structure which is seen around W4 is generated by Faraday rotation
occurring in ionized gas near the \ion{H}{ii} regions \citep{gray99};
the same may apply to the Arc. We note that the apparent centre of
curvature of the Arc is at ${\ell}{\approx}{135^{\circ}},
{b}{\approx}{8^{\circ}}$, at virtually the same longitude as W4. The
Arc may have been flattened by the density gradient below W4
\citep{norm96}, in which case the latitude of its centre is lower,
within the W4 complex. We suggest that the W4 Polarization Arc may be
at the same distance as the \ion{H}{ii} complex, and it may be related
to the star-formation activity there.

\citet{norm96} showed that the stars which ionize W4 (the star cluster
OCl~352) have, through the influence of their winds, blown a large
bubble above W4 (to positive Galactic latitudes). Although originally
thought to be a Galactic ``chimney'' (a conduit for hot gas and
ionizing radiation from the plane of the Galaxy into the halo),
subsequent observations \citep{denn97,west07} have shown the bubble is
nearly closed about 325 pc above the plane. \citet{reyn01} presented
H${\alpha}$ data which reveal an ionized shell 1.3 kpc above W4. This
ionized material, at the level where the Galactic disk blends into the
halo, is probably a feature of an earlier phase of star formation in
the same vicinity (discussed in West et al. 2007). The Polarization
Arc can be accommodated in a scenario where successive generations of
star formation have made an impact over hundreds of parsecs of the
ISM.

\subsection{The planetary nebula Sharpless 2-216}
\label{subsec:s216}

The planetary nebula Sharpless 2-216 generates a polarization feature
(at ${\ell}{\approx}{158^{\circ}}$, ${b}{\approx}{1^{\circ}}$) through
Faraday rotation in its ionized shell \citep{rans08}. Sh 2-216 is the
closest planetary nebula, at a well-established distance of
129~pc. Its angular size of 1\fdg7 translates to a physical diameter
of 3.8~pc. The ionized shell, of density $\sim$10 cm$^{-3}$, generates
a weak feature in total intensity but a strong Faraday rotation
signature with ${{\Delta}{\psi}}\approx{60^{\circ}}$. Electron density
in the shell can be deduced from optical and radio data, and path
length is precisely known: the magnetic field can therefore be
established. The line-of-sight field component is $\sim$5~$\mu$G,
probably the interstellar field, slightly enhanced by compression
in the planetary nebula shell. This is the first detection of magnetic
field in a planetary nebula.

\subsection{Stellar-wind bubble in the anticentre}
\label{subsec:anticentre_bubble}

A large arc in PI is evident at
${162^{\circ}}<{\ell}<{169^{\circ}}, {4^{\circ}}>{b}>{2^{\circ}}$
(see Fig.~\ref{ipi158176}).
Close examination shows that this is a circular structure, centred at
${\ell}={166^{\circ}}, {b}={-1.0^{\circ}}$. A search of the CGPS
database shows a clearly associated arc of \ion{H}{i} at
${v_{lsr}}={-17}${\thinspace}km{\thinspace}s$^{-1}$; we refer to the
object as GSH~166$-$01$-$17.  This association places
GSH~166$-$01$-$17 in the Perseus Arm at a distance of $\sim$1.9~kpc
where its extent is some 350 pc.

The interpretation (Kothes et al., in preparation) sees
GSH~166$-$01$-$17 as a superbubble blown by a cluster of stars over a
period of 10 to 20 million years.  A group of stars, ranging in type
from B2 to B4, is still within the shell but there must have been
stars of earlier type present to blow such a large bubble, and those
stars have probably now exploded as supernovae, contributing to the
internal energy of the bubble. There is (weak) evidence of non-thermal
emission within the bubble, possibly the result of supernova
explosions within the shell, and diffuse X-ray emission, presumably
from hot interior material, can be seen in the ROSAT X-ray
survey. Beyond these traces, the bubble sits in a remarkably empty
part of the Galaxy. However, the bubble is surrounded by young
objects, of typical age 1 million years, and it is conceivable that
this is star formation triggered by the evolution of GSH~166$-$01$-$17.

A simple interpretation, based on the model of \citet{weav77} can explain the polarization signature of the
stellar-wind bubble. In the Weaver et al. model the centre of such a bubble is filled
with very hot gas (temperature ${\sim}10^6$~K), of very low
density, surrounded by a thick shell of shocked stellar-wind material at a
temperature of ${\sim}10^5$~K. In GSH~166$-$01$-$17 there is apparently a toroidal magnetic field within
this shell. Background polarized emission is Faraday rotated
as it passes through the shell and adds vectorially to a polarized
foreground.  PI rises and falls, as seen in Fig.~\ref{ipi158176}, as
the rotated background and the foreground components reinforce and
cancel.  The toroidal field is presumably stellar in origin. If this interpretation
is correct, GSH~166$-$01$-$17 is a prime example
of stellar winds carrying magnetic energy from stars into the ISM.

Bubbles and superbubbles play a prominent role in current models of
ISM evolution \citep{deav05}, and these objects are widely supposed to
be affected by magnetic fields \citep{ferr91,tomi98,stil09}.
Nevertheless, there are very few measurements of magnetic fields in
such objects (for another example see West et al. 2007).

\subsection{The high-latitude region}
\label{subsec:highlat}

\begin{figure*}
%\centerline{\includegraphics[bb = 10 12 555 660,width=11.8cm,angle=-90,clip]
%    {f9.eps}}     
%\resizebox*{11.8cm}{!}{\centerline{\includegraphics[angle=-90]{f9}}}
\resizebox*{17cm}{!}{\includegraphics[angle=-90]{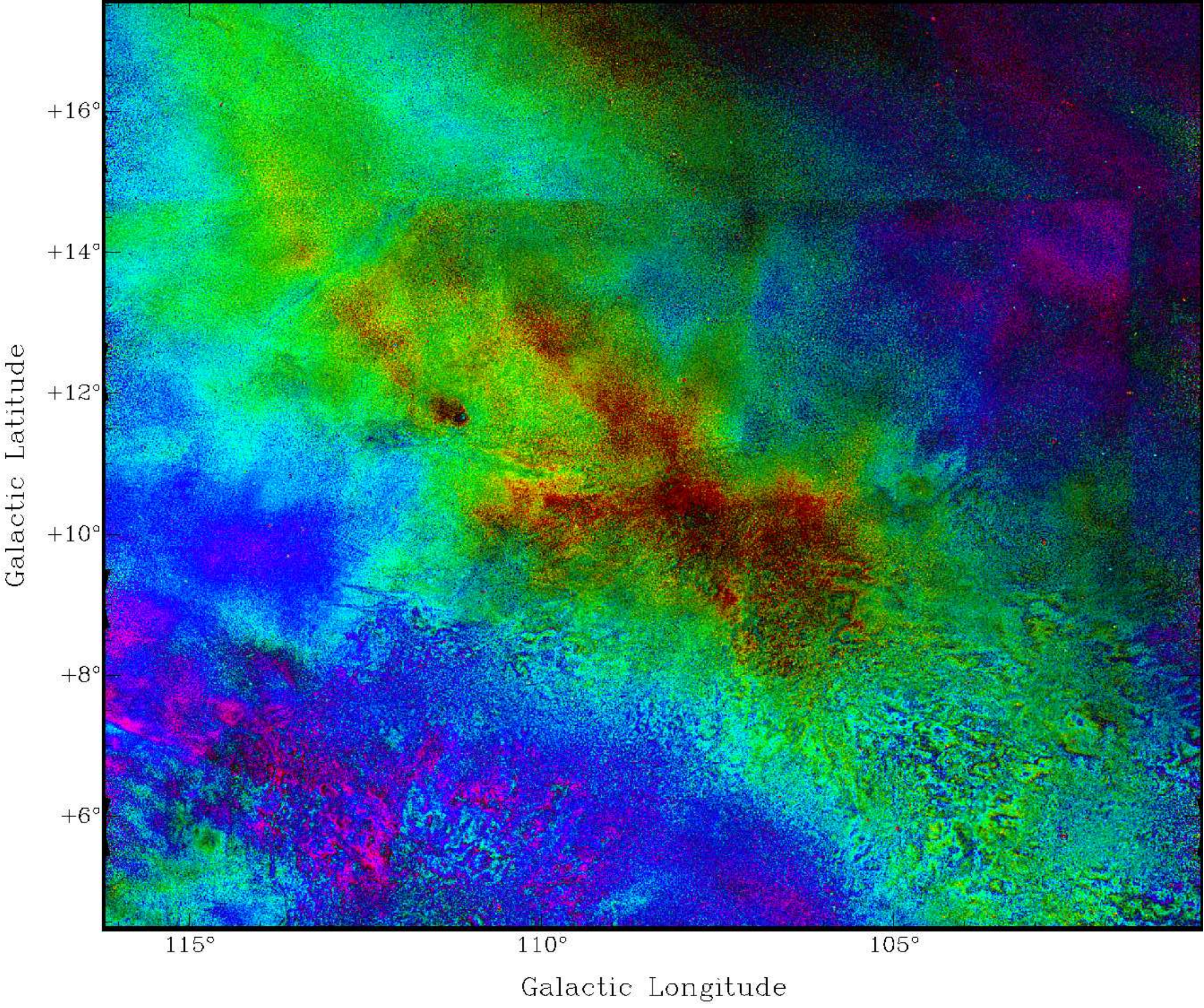}}
   \caption
{Polarization data in the high-latitude region. As in
Fig.~\ref{whole_survey}, intensity of colour depicts polarized
intensity and hue depicts polarization angle. The range of polarized
intensity is zero (dark) to 0.5~K (bright). The range of polarization
angle is from $-40^{\circ}$ to $+10^{\circ}$. Colour changes smoothly
from red (${\rm{PA}}{\approx}{-40^{\circ}}$) through orange, yellow,
green, blue, and purple, to red (${\rm{PA}}{\approx}{+10^{\circ}}$).}
   \label{highlat}
\end{figure*}

\begin{figure*}
%\centerline{\includegraphics[bb = 203 70 463 687,width=7cm,angle=-90,clip]
%    {f10.eps}}     
\resizebox*{18cm}{!}{\includegraphics[angle=-90]{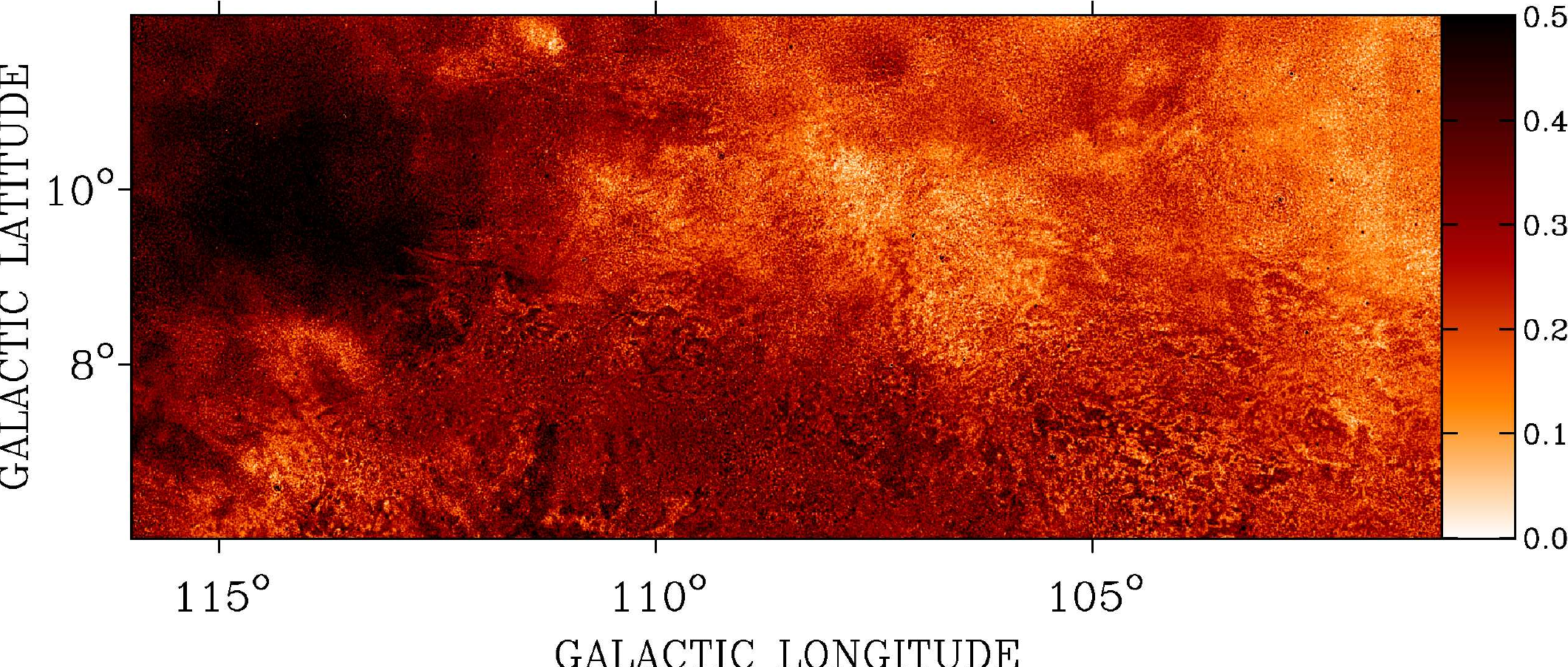}}
   \caption
{Polarized intensity in the high-latitude region, showing a transition
in structure between ${b}\approx{8^{\circ}}$ and ${b}\approx{10\fdg5}$.}
   \label{diskhalo}
\end{figure*}

The high-latitude region, ${101^{\circ}}\leq{\ell}\leq{116^{\circ}}$,
${4\fdg5}\leq{b}\leq{16\fdg5}$ is shown in Fig.~\ref{highlat}. A
rectangular boundary can be seen with edges at
${\ell}\approx{101\fdg5}$ and ${b}\approx{14\fdg5}$.  Effelsberg data
were not included outside these limits (to the upper and right hand edges
of Fig.~\ref{highlat}).

A striking feature of Fig.~\ref{highlat} is a transition from
small-scale structure at lower latitudes to a much smoother structure
above it. This transition is seen more clearly in
Fig.~\ref{diskhalo} which shows PI only. The transition begins at
${b}{\approx}{8^{\circ}}$ and is complete across most of the longitude
range shown by ${b}{\approx}{10\fdg5}$. Below the transition the size
of structures is about $3'$, but above it structure is distinctly
smoother with scale size at least $20'$. The transition does not
appear to be related to the change in PA that runs diagonally across
Fig.~\ref{highlat} (seen as changing colour); that is probably a
foreground effect.  Such a transition in structure is suggestive of an
interface between disk and halo regions. The very fact that we detect
an abrupt transition implies that the boundary that we are seeing
cannot be a feature of the Local Arm (we would not see a sharp
transition from within the disk) -- we must be seeing the ``top'' of
the Perseus Arm, at a distance of $\sim$3~kpc. At this distance the
transition at ${b}\approx{10^{\circ}}$ is at a physical distance
$\sim$500~pc above the Galactic mid-plane. The characteristic sizes of
$3'$ and $20'$ below and above the transition correspond to physical
sizes of 3~pc and 17~pc respectively. The transition is reminiscent of
the model of \citet{kalb98} who envisage an abrupt transition between
a turbulent disk and a smooth halo in which gas, cosmic rays and
magnetic fields are in pressure equilibrium.

\begin{figure}
%   \centerline{\includegraphics[bb = 35 65 540 375,width=9cm,clip]
%    {f11.eps}}     
\resizebox*{9cm}{!}{\includegraphics[angle=0]{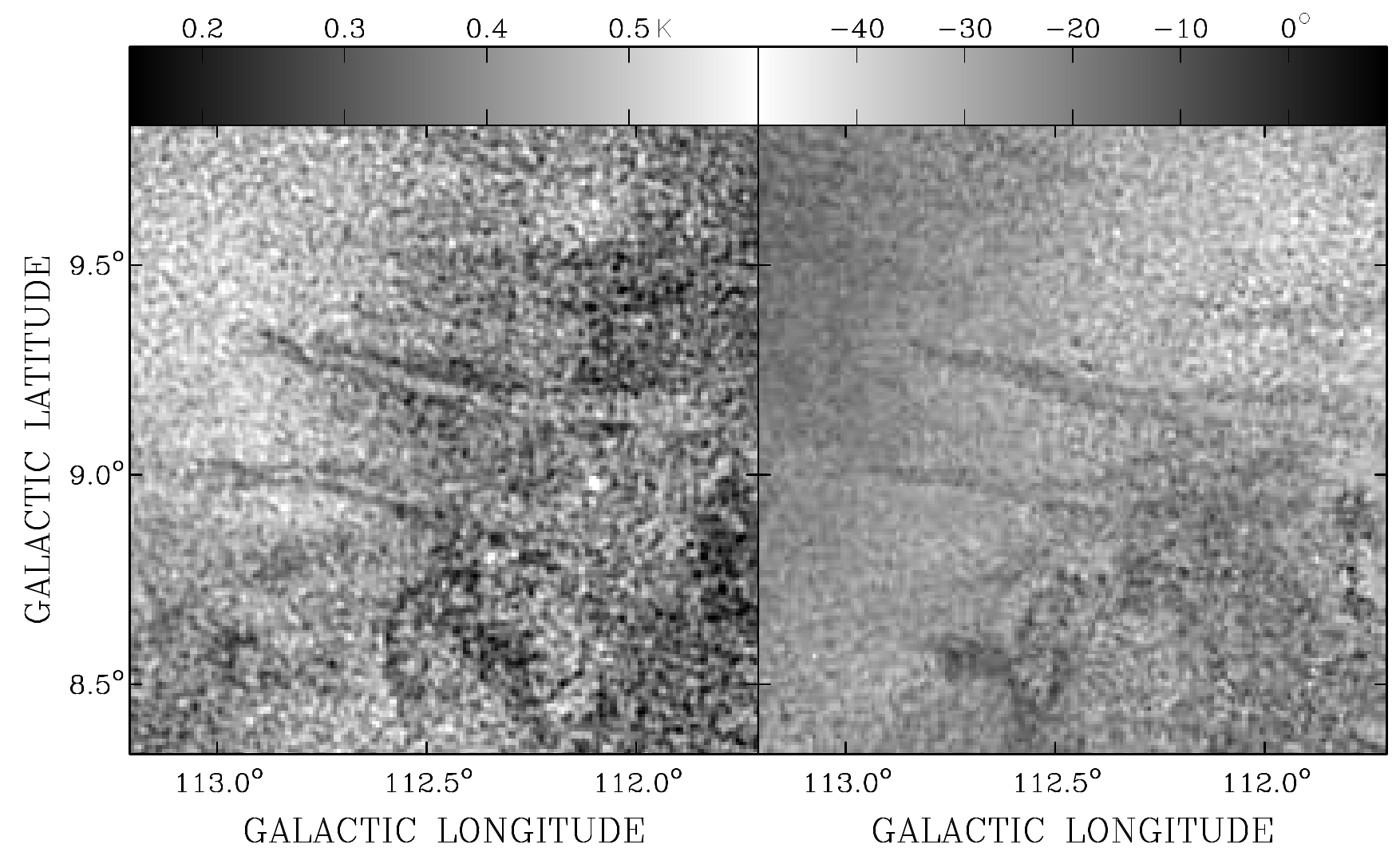}}
   \caption {The ``fingers'' at the interface between disk and halo
    regions.  left panel shows polarized intensity and right panel
    polarization angle.  See Section~\ref{subsec:highlat}.}
   \label{fingers}
\end{figure}

Figure~\ref{fingers} is enlarged to show one detail of the disk-halo
interface. Three long ``fingers'' are seen in PI and PA, about $5'$
wide and $45'$ long. At a distance of 3~kpc they are $\sim$480~pc
above the mid-plane and their physical dimensions are 4~pc wide and
40~pc long. These fingers may be structures in magnetic field,
structures in electron density, or some combination of the
two. The change of ${\sim}20^{\circ}$ that they produce in
polarization angle implies a RM of about 7
rad{\thinspace}m$^{-2}$. The fingers cannot be detected in available
H{\thinspace}${\alpha}$ images and there is no other information on
electron density, so the magnetic field within them cannot be
deduced. They are suggestive of some kind of magnetohydrodynamic
instability at an interface between two fluids.

\section{Conclusions, and extensions of this work}
\label{sec:conclus}

We have described techniques developed to make a survey of the
polarized radio emission from the Galactic plane over a large area,
combining data from aperture-synthesis and single-antenna telescopes
to provide an accurate portrayal of emission features on all angular
scales to the resolution limit of ${\sim}1'$.  This survey represents
a major advance in high-fidelity imaging of the polarized sky.  We
have presented data from the survey and have made preliminary
interpretations of some features revealed by it.

Mapping the polarized sky opens a new ``window'' on the ISM because
the appearance of the sky is dominated by Faraday rotation occurring
along the propagation path through the Galaxy, to the point where the
polarized sky does not resemble the total-intensity sky.  Some general
conclusions can be reached from this work that will be relevant to
future polarization imaging.

\begin{itemize}

\item {While Faraday rotation tends to break up large emission
structures into smaller ones, there is still significant large-scale
structure at 1.4~GHz. Consequently, interpretation of
aperture-synthesis data will be severely limited unless single-antenna
data are accurately incorporated into the images.}

\item{Faraday rotation is a powerful tool for detecting ionized gas, 
and polarization observations will lead to the discovery of objects 
that cannot easily be detected by other means.}

\item{There are features of the Galactic emission many degrees in
extent; despite their large size they are very difficult to recognize
without arcminute angular resolution.}

\item{Polarization features trace structures in the magneto-ionic
medium, in electron density or magnetic field or both, and
observations of the diffuse polarized emission reveal a diversity of
phenomena associated with this component of the ISM. The magneto-ionic
medium is in part unstructured and very broadly distributed but it may
also be associated with discrete objects such as SNRs, \ion{H}{ii}
regions, planetary nebulae, and stellar-wind bubbles.}

\item{Some of the features seen in polarization images are the
products of propagation and depolarization effects of various kinds,
and are not necessarily ``objects'' in the usual astronomical sense.}

\end{itemize}

This survey has revealed a wealth of structure in the magneto-ionic
medium on all scales. Future work on the data presented here will be
directed at illuminating the relationship between these structures and
other phases of the ISM. Information on other ISM tracers,
particularly the CGPS datasets describing the atomic, ionized, and
molecular gas and the dust, will be critical to the success of these
studies.

The longitude range of the survey has been extended to
${\ell}{\approx}{55^{\circ}}$ towards the inner Galaxy and to
${\ell}{\approx}{195^{\circ}}$ beyond the anticentre.  These new data
will be processed using the techniques described in this paper.

\begin{acknowledgements}
The Dominion Radio Astrophysical Observatory is operated as a national
facility by the National Research Council Canada. The Canadian
Galactic Plane Survey is a Canadian project with international
partners. The survey was supported by a grant from the Natural Sciences
and Engineering Research Council of Canada. This research is in part
based on observations with the 100-m Telescope of the
Max-Planck-Institut f\"ur Radioastronomie at Effelsberg. The
polarization survey using the DRAO 26-m Telescope was a joint project
between DRAO and MPIfR.  

It is a pleasure to acknowledge the outstanding and dedicated work of
Diane Parchomchuk, Jack Dawson, Ev Sheehan, Rod Stewart, Jean Bastien
and Tony Hoffmann in operating and maintaining the DRAO Synthesis and
26-m Telescopes, in assessing data, diagnosing telescope faults, and
fixing them. We could not have completed this work without the
exceptional image processing algorithms developed by Lloyd Higgs and
Tony Willis. We thank Rainer Beck for valuable comments on the paper.

\end{acknowledgements}

\bibliographystyle{aa}
%\bibliography{cgpspolsurvey_rev}
\bibliography{preprint_pdf}

\end{document}